\documentclass[a4paper, 12pt]{article}

\pdfoutput = 1

\usepackage{styleBuding}
\usepackage{bm}
\usepackage{color}
\usepackage[usenames,dvipsnames,svgnames,table]{xcolor}
\usepackage{amsmath}
\usepackage{amssymb}
\usepackage{graphicx}
\usepackage{slashed}
\usepackage{soul}
\usepackage{multirow}
\usepackage{subfigure}
\usepackage{siunitx}
\usepackage{url}
\usepackage{placeins}
\usepackage[utf8]{inputenc}

\newcommand{\rom}[1]{\uppercase\expandafter{\romannumeral #1\relax}}

\allowdisplaybreaks
\DeclareUnicodeCharacter{2212}{-}

\let\jnfont=\rm
\def\NPB#1,{{\jnfont Nucl.\ Phys.\ B }{\bf #1},}
\def\PLB#1,{{\jnfont Phys.\ Lett.\ B }{\bf #1},}
\def\EPJC#1,{{\jnfont Eur.\ Phys.\ Jour.\ C }{\bf #1},}
\def\PRD#1,{{\jnfont Phys.\ Rev.\ D }{\bf #1},}
\def\PRL#1,{{\jnfont Phys.\ Rev.\ Lett.\ }{\bf #1},}
\def\MPLA#1,{{\jnfont Mod.\ Phys.\ Lett.\ A }{\bf #1},}
\def\JPG#1,{{\jnfont J.\ Phys.\ G}{\bf #1},}
\def\CTP#1,{{\jnfont Commun.\ Theor.\ Phys.\ }{\bf #1},}
\def\ZPC#1,{{\jnfont Z.\ Phys.\ C }{\bf #1},}
\def\JHEP#1,{{\jnfont JHEP \ }{\bf #1},}

\title{Probing LZ-Motivated TeV Higgsino Dark Matter through Heavier Sparticles at LHC}

\author{Xintian Wang$^{a}$,~Jin Min Yang$^{a,b}$,~Yuanfang Yue$^{a}$,~Yang Zhang$^{a}$}

\affiliation[a]{School of Physics, Henan Normal University,
Xinxiang 453007, China}
\affiliation[b]{Institute of Theoretical Physics, Chinese Academy of Sciences, Beijing 100190, China}

\emailAdd{107552300777@stu.xju.edu.cn}
\emailAdd{jmyang@itp.ac.cn}
\emailAdd{yueyuanfang@htu.edu.cn}
\emailAdd{zhangyang2025@htu.edu.cn}

\date{September 2026}

\abstract{
The nearly pure TeV-scale Higgsino dark matter provides a natural explanation for the high-energy nuclear-recoil excess reported by LUX-ZEPLIN.
Such a heavy and compressed Higgsino spectrum makes direct searches at the LHC challenging.
We therefore investigate complementary probes through the production and decay of heavier sparticles.
We perform a comprehensive analysis and obtain the following observations:  
(i) For a TeV Higgsino, the current LHC Run-2 searches exclude gluino up to approximately 2.2~TeV, while 
the HL-LHC can give an exclusion reach of 2.5--2.6~TeV and a discovery reach of 2.3--2.4~TeV;
(ii) The light-flavor squark coannihilation window lies beyond the Run-2 reach, but its lower end could be excluded at the HL-LHC under optimistic assumptions;
(iii) The Run-2 search does not exclude the stop or sbottom benchmarks;
(iv)  The slepton and heavier-electroweakino productions are too small to access at the HL-LHC.
}

\begin{document}

\maketitle

\section{Introduction}
\label{sec:introduction}

The LUX-ZEPLIN (LZ) experiment recently extended its nuclear-recoil
search window to approximately $270~{\rm keV}$ using an exposure of
$2.84$ tonne-years. It observed one event consistent with a nuclear
recoil of $248\pm23\,{\rm (stat)}\pm23\,{\rm (sys)}~{\rm keV}$ in a
region of low expected background. The background-only hypothesis has
a global significance of $2.6\sigma$ after the look-elsewhere effect,
with a maximum local significance of $3.4\sigma$ among the interaction
models tested. LZ examined rare background processes and detector
effects but identified none as a likely explanation of the event
\cite{LZ:2026axp}. Its unusually high recoil energy opens a qualitatively
new window for dark-matter interactions whose signals are suppressed at
low recoil energy and enhanced at large momentum transfer.

The 248~keV recoil has prompted a broad range of interpretations. Studies
of Higgsinos and other electroweak multiplets connect the recoil energy
to the relic abundance, the origin of the sub-MeV splitting, and the
elastic loop amplitude
\cite{Yin:2026jnn,Nomura:2026qyq,DiMauro:2026ldr,Visinelli:2026kgt,
Chattopadhyay:2026ryw,Smirnov:2026aqk,Du:2026guj,Wang:2026ytg,
Bandyopadhyay:2026gjw,Borah:2026zwf,Bisal:2026khf,
Langhoff:2026ujr,Chatterjee:2026scv,Frolovsky:2026tvq,Lian:2026hpm,
Girmohanta:2026qzo,Khan:2026zuj}.
Broader particle constructions use dark photons, axion-like or PQ
portals, additional gauge symmetries, magnetic transitions,
neutrino-mass sectors, and nonstandard production histories
\cite{Yamashita:2026ump,Unwin:2026rdp,Lee:2026wof,
Ahmed:2026qjg,Yuan:2026djt,Zhu:2026dag,Asadi:2026iot,Lee:2026xxh,
Lee:2026jxl,He:2026hqz,Qi:2026vyp,Okada:2026upm,Borah:2026ris,Barman:2026omh,He:2026idw,An:2026pkc,Xing:2026civ,Nagata:2026pbj,
Okada:2026fef}.
Other analyses examine exothermic or boosted particles, dark-matter
absorption, neutrino scattering, xenon excitation, and neutron
disappearance
\cite{Lou:2026idn,Jeesun:2026vzo,deLima:2026shq,Gu:2026vto,
Baer:2026fpy,Liang:2026coz,Alhazmi:2026efz,Kannike:2026qyl,
Khan:2026nwp,Heikinheimo:2026kwp,
Uttayarat:2026isp,Fan:2026hzw,
Mahapatra:2026glu,Palmisano:2026kuj}. The empty high-energy sideband,
annual modulation, solar-neutrino and gamma-ray searches, and
complementary detector targets can separate these mechanisms
\cite{Su:2026rwz,Pospelov:2026ewn,Rodd:2026tyn,McCabe:2026crm,
Dent:2026bji,Yang:2026wpb,DiMauro:2026dqp,Bose:2026ndd,Li:2026fci,
Nguyen:2026lui,Ghosh:2026txe,Baer:2026yrt,
DiMauro:2026ymt}.

Among these possibilities, we focus on the nearly pure Higgsino
interpretation for its predictive power. If the Higgsino constitutes
all dark matter and colored coannihilators are heavy, thermal freeze-out
fixes its mass near $1.1~{\rm TeV}$. The off-diagonal $Z$ interaction
fixes the inelastic scattering strength. The LZ recoil then selects mainly the
neutral-state splitting rather than an arbitrary cross section
\cite{Freese:2026sga,Wu:2026nhi}. The same electroweak interaction
captures Higgsinos in the Sun, where their annihilation into
$W^+W^-$ and $ZZ$ produces high-energy
neutrinos~\cite{Luengas:2026tlf}. If the captured population
thermalizes efficiently, ten years of IceCube data require
a splitting of about $0.55~{\rm MeV}$ or larger, which disfavors the
standard-halo interpretation of the LZ event
\cite{Pospelov:2026ewn,Bose:2026ndd}. The high-velocity tail generated
by the Large Magellanic Cloud(LMC)~\cite{Bozorgnia:2026emp} shifts the
LZ-preferred splitting to
$481$--$519~{\rm keV}$. Near an elastic-scattering blind spot, the
captured Higgsinos remain nonthermal and the IceCube lower bound falls
to about $475~{\rm keV}$, leaving the LMC-favored region viable
\cite{Wu:2026nhi}.

The LZ-motivated Higgsino spectrum is difficult to probe directly at the
LHC because electroweak production is small and the compressed multiplet
produces only soft decay products\cite{Maier:2026qpr}. HL-LHC monojet searches with soft
leptons reach about $200~{\rm GeV}$~\cite{Schwaller:2013baa}, while
disappearing-track searches require sub-centimeter tracklet
reconstruction to approach the $1.1~{\rm TeV}$ thermal target
\cite{Cheung:2026byg}. A $100~{\rm TeV}$ collider can cover this region
by combining central and forward disappearing-track selections
\cite{Mahbubani:2017gjh}. At muon colliders, a $10~{\rm TeV}$ machine
can discover the thermal Higgsino with about $8~{\rm ab}^{-1}$ through
disappearing tracks~\cite{Capdevilla:2021fmj}, while a $3~{\rm TeV}$
machine can use a hard photon to trigger events containing soft charged
decay products~\cite{Capdevilla:2024bwt}. The direct reach is therefore
set by the collider energy, the innermost tracking radius, and
low-momentum reconstruction
\cite{Low:2014cba,Han:2018wus,Bhattiprolu:2025zwt,Cheung:2026byg}.

Although direct searches face these limitations, the LZ recoil does not
require the rest of the supersymmetric spectrum to decouple. It
constrains the bino and wino only through the neutral Higgsino splitting,
which can arise from heavy gauginos or a cancellation between
opposite-sign contributions. The recoil otherwise places no comparable
constraint on the superpartner spectrum beyond the mass ordering required
for a Higgsino LSP. If heavier superpartners lie near the Higgsino scale,
their production and decay at the LHC can provide hard visible objects and
missing momentum. We quantify the resulting collider sensitivity using
representative spectra of heavier superpartners.

The remainder of this paper is organized as follows.
Section~\ref{sec:benchmark} determines the Higgsino spectrum from the
relic density and the LZ recoil. Section~\ref{sec:collider_probes}
studies its LHC signatures from heavier superpartners.
Section~\ref{sec:conclusions} summarizes our results.

\section{The LZ-motivated Higgsino scenario}
\label{sec:benchmark}

The LZ interpretation selects a nearly pure Higgsino multiplet with two
neutral states $\tilde\chi^0_{1,2}$ and one charged state,
$\tilde\chi_1^\pm$. Their common mass is set predominantly by the Higgsino mass parameter $\mu$.
We assume that the lightest neutralino constitutes all dark matter.
With colored superpartners decoupled, electroweak annihilation and
coannihilation fix the Higgsino mass
through the relic abundance. Requiring agreement with the Planck
measurement~\cite{Planck:2018vyg}, while allowing a $10\%$ theory
uncertainty, gives
$m_{\tilde\chi_1^0}\simeq1.09~{\rm TeV}$, with ranges of
$1.00$--$1.18~{\rm TeV}$ at $1\sigma$ and
$0.94$--$1.23~{\rm TeV}$ at $2\sigma$, which are obtained using
\textsc{micrOMEGAs}~7.1~\cite{Belanger:2026asz,Alguero:2023zol}. 

To connect the LZ recoil~\cite{LZ:2026axp} to the Higgsino spectrum, we
determine the neutral-state splitting required by inelastic scattering.
For a nearly pure Higgsino, the neutral current is predominantly off
diagonal, while the weak interaction fixes the Higgsino--nucleon cross
section at
$\sigma_{\tilde H N}\simeq1.86\times10^{-39}~{\rm cm}^2$.
Consequently, the observed recoil energy constrains the inelastic
threshold rather than an adjustable interaction strength. At recoil
energy $E_R$, the minimum incident speed is
\begin{equation}
    v_{\min}(E_R)
    =\frac{1}{\sqrt{2m_AE_R}}
     \left(\frac{m_AE_R}{\mu_{\chi A}}+\delta_0\right),
    \qquad
    \delta_0=m_{\tilde\chi_2^0}-m_{\tilde\chi_1^0}.
    \label{eq:inelastic_vmin}
\end{equation}
Here, $m_A$ denotes the nuclear mass and $\mu_{\chi A}$ the
dark-matter--nucleus reduced mass. The inferred splitting depends on
the high-speed tail of the halo
distribution~\cite{OHare:2026nqi,Das:2026jsq}. Assuming the standard
halo model, a fit to the
$248\pm23\,{\rm(stat)}\pm23\,{\rm(sys)}~{\rm keV}$ event peaks at
$\delta_0=366~{\rm keV}$ and gives $312$--$422~{\rm keV}$ at $1\sigma$
and $280$--$457~{\rm keV}$ at $2\sigma$. Including an LMC-enhanced
high-velocity tail shifts the favored splitting to
$481$--$519~{\rm keV}$ over the $1\sigma$ mass interval
\cite{Freese:2026sga,Wu:2026nhi}. We adopt the representative
standard-halo point
\begin{equation}
    m_{\tilde\chi_1^0}=1.10~{\rm TeV},
    \qquad
    \delta_0=350~{\rm keV}
    \label{eq:higgsino_benchmark}
\end{equation}
for the following collider calculation. Both the standard-halo and
LMC-favored splittings are sub-MeV, so this choice does not affect the
collider kinematics studied below.

The same neutral splitting fixes the lifetime of the excited state.
For $\delta_0$ below a GeV, the dominant decay is the radiative process
$\tilde\chi_2^0\to\tilde\chi_1^0\gamma$ induced by a $W$ loop. At the
thermal benchmark its proper decay length is approximately
\begin{equation}
    c\tau_{\tilde\chi_2^0}
    \simeq 8.0\times10^3~{\rm km}
    \left(\frac{350~{\rm keV}}{\delta_0}\right)^3,
    \label{eq:neutral_higgsino_lifetime}
\end{equation}
where the $\mu$ dependence is weak over the relic-density
interval~\cite{Fan:2026kxx}. At $|\mu|=1.1~{\rm TeV}$, the
standard-halo $2\sigma$ range gives
$c\tau\simeq(3.6\times10^3$--$1.6\times10^4)~{\rm km}$ and
$\tau\simeq0.012$--$0.052~{\rm s}$. The LMC range gives
$c\tau\simeq(2.5$--$3.1)\times10^3~{\rm km}$ and
$\tau\simeq0.0082$--$0.010~{\rm s}$. This lifetime is invisible at
colliders. 

The charged-state splitting follows from the same spectrum but is
dominated by electroweak radiative corrections.  In the heavy-gaugino
limit its one-loop contribution is
\begin{equation}
    \Delta m_\pm^{\rm rad}
    =\frac{\alpha_2}{4\pi}m_{\tilde\chi_1^\pm}s_W^2
      f\!\left(\frac{m_Z}{m_{\tilde\chi_1^\pm}}\right),
    \quad
    f(x)=2\int_0^1\!dt\,(1+t)
      \ln\!\left[1+\frac{x^2(1-t)}{t^2}\right]
    \label{eq:charged_loop_splitting}
\end{equation}
\cite{Nagata:2014wma}.  For the same-sign decoupling realization
$M_1=M_2>0$, positive $\mu$, and $\tan\beta=10$, the leading tree
contribution is
\begin{equation}
    \Delta m_\pm^{\rm tree}
    \simeq\frac{\delta_0}{2}
    \left(1-\sin2\beta\cos2\theta_W\right).
    \label{eq:charged_tree_splitting}
\end{equation}
Numerical evaluation gives
$\Delta m_\pm^{\rm rad}=343.046~{\rm MeV}$ and
$\Delta m_\pm^{\rm tree}=0.156~{\rm MeV}$ at the benchmark, hence
\begin{equation}
    \Delta m_\pm
    \equiv m_{\tilde\chi_1^\pm}-m_{\tilde\chi_1^0}
    =343.203~{\rm MeV}.
    \label{eq:charged_splitting}
\end{equation}
Scanning the relic-density and LZ $2\sigma$ ranges gives
$341.10<\Delta m_\pm/{\rm MeV}<344.59$.  The corresponding chargino
decays mainly through
$\tilde\chi_1^\pm\to\tilde\chi_1^0\pi^\pm$, with
$c\tau\simeq0.79$--$0.82~{\rm cm}$, or
$\tau\simeq0.026$--$0.027~{\rm ns}$, and a soft pion. Three- or
four-hit pixel tracklets target the disappearing chargino. A
low-momentum displaced-pion search provides a complementary signature,
with an initial-state-radiation jet supplying the trigger and missing
transverse momentum. ATLAS excludes pure-Higgsino charginos up to about
$225~{\rm GeV}$ for lifetimes below $0.03~{\rm ns}$ using
$137~{\rm fb}^{-1}$ of Run-2 data~\cite{ATLAS:2026hnb}. The CMS
soft-track search excludes $0.28<\Delta m_\pm/{\rm GeV}<1.15$ and
chargino masses up to
$185~{\rm GeV}$~\cite{CMS:2026ias}.  The splitting and lifetime predicted
here lie directly in the targeted region, but the $1.1~{\rm TeV}$
thermal Higgsino remains well beyond the current LHC reach.

The same small neutral splitting constrains the bino and wino masses.
Away from a gaugino--Higgsino crossing, the leading finite-mass
expression is
\begin{equation}
\begin{split}
    \delta_0^{\rm tree}&\simeq m_Z^2|C_B+C_W|,\\
    C_B&=s_W^2\frac{M_1+\mu\sin2\beta}{M_1^2-\mu^2},
    \qquad
    C_W=c_W^2\frac{M_2+\mu\sin2\beta}{M_2^2-\mu^2}.
\end{split}
    \label{eq:neutral_splitting_finite_gaugino}
\end{equation}
For $|M_{1,2}|\gg|\mu|,m_Z$, this reduces to
\begin{equation}
    \delta_0
    \simeq
    m_Z^2
    \left|
        \frac{s_W^2}{M_1}
        +
        \frac{c_W^2}{M_2}
    \right|,
    \label{eq:neutral_splitting_heavy_gaugino}
\end{equation}
up to corrections suppressed by the electroweakino mass ratios
\cite{Nagata:2014wma,Freese:2026sga,Wu:2026nhi}. Contributions with
the same sign add constructively and require both gauginos to decouple,
whereas opposite signs permit a cancellation at finite masses~\cite{Roy:2026wzb}.

For $M_1=M_2=M>0$, the LZ-preferred splitting requires
$M=\mathcal O(10^4)~{\rm TeV}$.
For opposite signs, the cancellation permits electroweak gauginos at
the TeV scale. We scan the full spectrum with \textsc{SPheno} and
\textsc{micrOMEGAs}~7.1, requiring the LZ splitting, a Higgsino-dominated
LSP, the observed relic abundance, and current direct-detection limits.
Spin-independent scattering removes the low-mass solutions and gives
$|M_1|\gtrsim1.68~{\rm TeV}$ and $M_2\gtrsim2.30~{\rm TeV}$.
The LZ recoil and direct-detection limits mainly constrain the
electroweakino sector. Satisfying the IceCube limit requires a
correlation among superpartner masses that cancels the tree-level and
loop-induced elastic amplitudes~\cite{Pospelov:2026ewn,Wu:2026nhi}, but
this correlation does not require the remaining superpartners to be
heavy. Consequently, relatively light states outside the electroweakino
sector can coexist with the LZ interpretation and provide additional
direct-production channels at the LHC.

\section{LHC probe through heavier sparticles}
\label{sec:collider_probes}

The compressed Higgsino multiplet is difficult to observe directly, but a heavier superpartner can make the same Higgsino visible by
supplying hard visible objects in its decay. The reach then depends on the
parent production rate, the mass gap, and the decay pattern. Throughout this section,
$m_{\widetilde H}$ denotes the common compressed-Higgsino mass scale and is
identified with $m_{\tilde\chi_1^0}$ in the decoupled-gaugino benchmarks.
We generate hard events using \textsc{MadGraph5\_aMC@NLO}
\cite{Alwall:2014hca}, showered with \textsc{Pythia}~8
\cite{Sjostrand:2014zea,Bierlich:2022pfr}, and passed through
\textsc{Delphes}~3\cite{deFavereau:2013fsa} and \textsc{CheckMATE}
\cite{Drees:2013wra,Dercks:2016npn}. Gluino and stop
branching fractions are obtained with \textsc{SDECAY}
\cite{Muhlleitner:2003vg}.

\subsection{Gluinos}
\label{sec:gluino}

Gluinos provide the most favorable production rate among the heavy parents
considered here. In this benchmark, only the gluino and Higgsino are light.
The other superpartners are decoupled, therefore the gluino decays through an
off-shell squark.
After the squark is integrated out, the three-body width scales as
\begin{equation}
    \Gamma(\tilde g\to q\bar q\tilde H)
    \propto
    \alpha_s y_q^2\frac{m_{\tilde g}^5}{m_{\tilde q}^4}
    \times\text{phase space}.
    \label{eq:gluino_width_scaling}
\end{equation}
Here $\alpha_s$ is the strong coupling and $y_q$ is the Yukawa coupling
of quark $q$.
The top Yukawa coupling then favors the heavy-flavor modes
\begin{equation}
    \tilde g\to t\bar t\,\tilde\chi^0_{1,2},
    \qquad
    \tilde g\to t\bar b\,\tilde\chi_1^-
    +\bar t b\,\tilde\chi_1^+ ,
    \label{eq:gluino_decays}
\end{equation}
with smaller $b\bar b\tilde\chi^0_{1,2}$ and
$g\tilde\chi^0_{1,2}$ contributions. At
$(m_{\tilde g},m_{\widetilde H})=(2.4,1.1)~{\rm TeV}$,
the summed $t\bar t\,\tilde\chi^0_{1,2}$ modes account for $45.25\%$
of gluino decays, while the $tb\,\tilde\chi_1^\pm$ modes account for
$51.28\%$. Gluino pairs therefore predominantly produce final states
with two, three, or four top quarks. The pure four-top topology accounts
for only about $(0.4525)^2\simeq20\%$ of events. The large
parent--daughter mass gap nevertheless gives hard jets and missing
momentum across this decay mixture, allowing it to be tested with an
inclusive hadronic search.

We test this signal with the CMS $M_{T2}$ search, which uses
$137~{\rm fb}^{-1}$ of $13$~TeV data and separates events by jet and
heavy-flavor content~\cite{CMS:2019ybf}. The signal is normalized to the
approximate NNLO+NNLL gluino-pair cross section. At each mass point we use
the published covariance block with the stronger expected sensitivity.
Appendix~\ref{app:gluino_validation} compares this recast with seven public $\tilde g\to t\bar t\tilde\chi_1^0$($T1tttt$) benchmarks\cite{CMS:2019ybf}. We characterize the
sensitivity by
\begin{equation}
    r_{\rm exp,obs}
    \equiv\frac{1}{\mu_{95}^{\rm exp,obs}},
    \label{eq:collider_rdef}
\end{equation}
where $\mu_{95}^{\rm exp,obs}$ is the expected or observed $95\%$ CL upper
limit on the signal strength. Thus $r_{\rm obs}>1$ denotes observed
exclusion, while $r_{\rm exp}>1$ denotes expected sensitivity at $95\%$ CL.

For the HL-LHC projection we simulate the signal at $14$~TeV and take
$\mathcal L=3~{\rm ab}^{-1}$. The energetic final states motivate using the
published aggregate signal regions with a Run-2-like detector response.
We estimate the $14$~TeV backgrounds in these signal regions by scaling
the Run-2 yields with luminosity and a rate factor $\kappa_B$. We take
$\kappa_B=1.0$, $1.2$, and $1.4$ to test how changes in the background
rate per unit luminosity affect the reach. We consider background
uncertainties of $10\%$, $20\%$, and $30\%$, as well as the Run-2 relative
total error. For the signal, we use the $15\%$ acceptance uncertainty
adopted in the CMS super signal region results~\cite{CMS:2019ybf}.

The exclusion band includes all of these background choices. For the
discovery band we use the fixed $10$--$30\%$ uncertainties. With the Run-2
relative total error, the simulated masses do not reach $5\sigma$.
We obtain the median expected exclusion and discovery sensitivities
from representative background-only and signal-plus-background data
sets respectively~\cite{Cowan:2010js,Read:2002hq}.

\begin{figure}[!htb]
    \centering
    \includegraphics[width=0.88\textwidth]
    {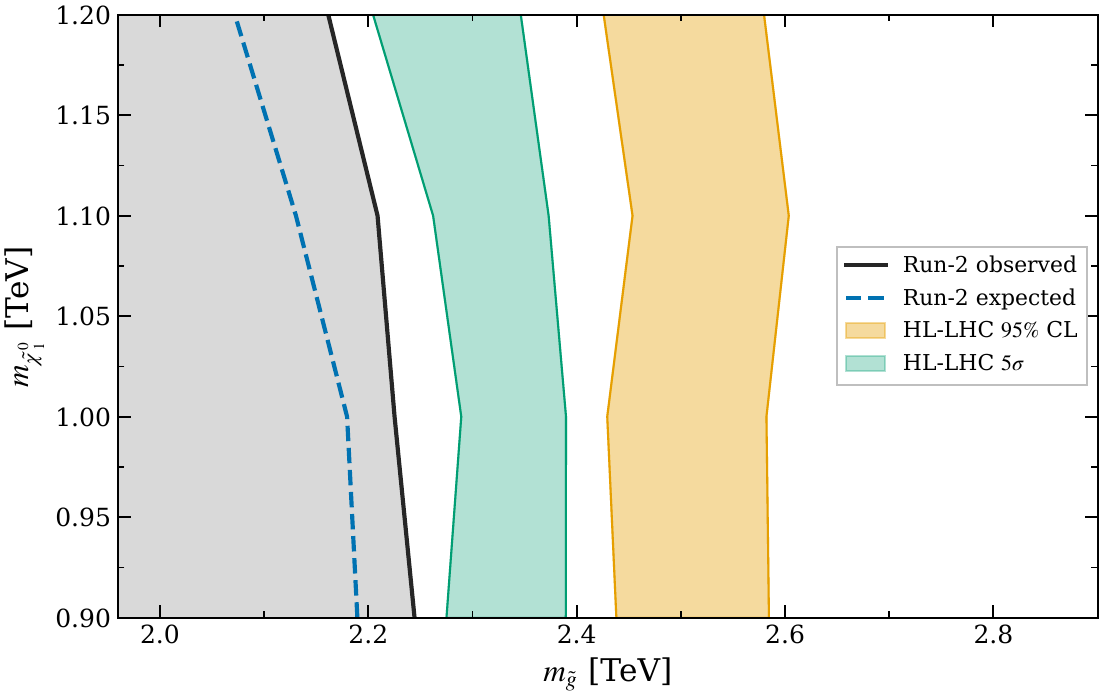}
    \caption{
    Run-2 gluino exclusion and projected HL-LHC sensitivity in the
    $(m_{\tilde g},m_{\tilde\chi^0_1})$ plane. The solid and dashed curves
    show the observed and expected Run-2 limits. The orange band covers
    expected $95\%$ CL exclusion across the background scenarios. The green
    band shows expected $5\sigma$ discovery for fixed background
    uncertainties. The shaded region to the left of the observed Run-2
    contour is excluded.
    }
    \label{fig:gluino_reach}
\end{figure}

Figure~\ref{fig:gluino_reach} displays the Run-2 boundaries and the HL-LHC
envelopes in the gluino--Higgsino mass plane. The solid and dashed curves
mark $r_{\rm obs}=1$ and $r_{\rm exp}=1$, while the two bands summarize the
HL-LHC exclusion and discovery projections defined above.
For the main $m_{\widetilde H}=1.1$~TeV benchmark, the expected and
observed Run-2 boundaries are
\begin{equation}
    m_{\tilde g}^{\rm exp}=2.13~{\rm TeV},
    \qquad
    m_{\tilde g}^{\rm obs}=2.21~{\rm TeV}.
    \label{eq:gluino_run2_main}
\end{equation}
The observed result thus excludes gluinos below about $2.21$~TeV for this
Higgsino mass.
The observed boundary changes only from about $2.24$ to $2.16~{\rm TeV}$
as the Higgsino mass increases from $0.9$ to $1.2~{\rm TeV}$. The large
remaining mass gap keeps the visible objects energetic, so the realistic
decay mixture retains substantial Run-2 sensitivity even though only a
minority of events have the pure four-top topology.

Accounting for the background-rate and uncertainty variations considered
above, we project the $95\%$ CL exclusion and $5\sigma$ discovery reaches
at the HL-LHC with an integrated luminosity of $3~{\rm ab}^{-1}$ to be
\begin{equation}
    m_{\tilde g}^{95\%\,{\rm CL}}
    =2.45\text{--}2.60~{\rm TeV},
    \qquad
    m_{\tilde g}^{5\sigma}
    =2.26\text{--}2.37~{\rm TeV}.
    \label{eq:gluino_hllhc_main}
\end{equation}
Once the mass gap is large enough to produce energetic heavy-flavor
objects, the reach is governed mainly by the steeply falling gluino
production rate. This explains the weak daughter-mass dependence in
Figure~\ref{fig:gluino_reach}. It also makes gluino production the strongest
quantified collider probe in this study, while the widths of the HL-LHC
bands show the remaining impact of the background uncertainty.

\FloatBarrier

\subsection{Squarks}
\label{sec:squarks}

\subsubsection{Light-flavor squarks}
\label{subsec:eight_squarks}

LHC searches for light-flavor squarks commonly use a degenerate benchmark
for the left- and right-handed states of the first two generations
\cite{ATLAS:2020syg}. This benchmark combines the QCD production of eight
squark states into a two-dimensional interpretation in the
$(m_{\tilde q},m_{\chi})$ plane. We therefore adopt the same degenerate
spectrum for our collider benchmark and recast validation,
\begin{equation}
    \tilde u_L,\ \tilde d_L,\ \tilde c_L,\ \tilde s_L,
    \qquad
    \tilde u_R,\ \tilde d_R,\ \tilde c_R,\ \tilde s_R.
    \label{eq:eight_light_squarks}
\end{equation}
Their flavor-diagonal decays produce light-flavor jets and missing
momentum. The mass gap to the
Higgsino is much smaller than in the gluino channel. We implement the
Higgsino couplings with \textsc{FeynRules} in
the Universal FeynRules Output format (UFO)
\cite{Alloul:2013bka,Degrande:2011ua}. Although the light-quark Yukawa
couplings make the widths small, they do not reduce the total branching
fraction to Higgsinos when no competing sparticle channel is open. A
partial width is
\begin{equation}
    \Gamma(\tilde q\to q\tilde\chi)
    \simeq\frac{|y_q|^2m_{\tilde q}}{16\pi}
    \left(1-\frac{m_{\tilde\chi}^2}{m_{\tilde q}^2}\right)^2,
    \label{eq:light_squark_width}
\end{equation}
Here $m_{\tilde\chi}$ is the daughter Higgsino mass. The appropriate
neutralino or chargino mixing factor has
been omitted. Even the least-coupled
$\tilde u_R$ has $c\tau\sim10^{-3}~{\rm cm}$ at
$(m_{\tilde q},m_{\widetilde H})=(1.15,1.10)~{\rm TeV}$. The benchmark
decays considered below are prompt.

For a right-handed squark, the pure-Higgsino limit gives equal summed
branching fractions to neutral and charged states,
\begin{equation}
    {\rm BR}(\tilde q_R\to q\tilde\chi^0_{1,2})
    \simeq {\rm BR}(\tilde q_R\to q'\tilde\chi_1^\pm)
    \simeq\frac12,
    \label{eq:right_light_squark_branching}
\end{equation}
with approximately one quarter in each neutralino. For left-handed
squarks the neutral fraction is set by the corresponding up- and
down-type Yukawa couplings. At $\tan\beta=10$, it is approximately
$(0.2\%,99.8\%)$ for $(\tilde u_L,\tilde d_L)$ and $(63\%,37\%)$ for
$(\tilde c_L,\tilde s_L)$. These branching differences have little effect
on the visible kinematics because the neutral and charged Higgsinos are
nearly degenerate. The jet momentum in the squark rest frame is therefore
controlled mainly by the parent--daughter mass gap,
\begin{equation}
    p_j^*=\frac{m_{\tilde q}^2-m_{\widetilde H}^2}{2m_{\tilde q}}.
    \label{eq:light_squark_jet_momenta}
\end{equation}
Close to threshold, the decay jet is soft and the signal relies strongly
on initial-state radiation. This motivates the compressed-spectrum search
used below.

The ATLAS multi-bin compressed (MB-C) selection targets compressed squark
spectra using energetic initial-state radiation~\cite{ATLAS:2020syg}.
It bins events by jet multiplicity, effective mass $m_{\rm eff}$, and
missing-momentum significance $E_T^{\rm miss}/\sqrt{H_T}$. Here $H_T$ is the
scalar jet-$p_T$ sum and
$m_{\rm eff}=H_T+E_T^{\rm miss}$. The search uses
$139~{\rm fb}^{-1}$ at $13$~TeV. Its public multibin likelihood is
evaluated through \textsc{Spey} with the \textsc{pyhf}
backend~\cite{Araz:2023bwx,Heinrich:2021gyp,Cranmer:2012sba}. We use the
sensitivity ratio defined in Eq.~(\ref{eq:collider_rdef}).
Appendix~\ref{app:squark_stop_validation} validates the MB-C recast against
the public degenerate $T2qq$ benchmark, in which each squark
decays to a quark and the lightest neutralino.

\begin{table}[htbp]
    \centering
    \caption{
    Rest-frame jet momenta and Run-2 MB-C sensitivity for degenerate
    light-flavor squarks. The sensitivity ratios are defined in
    Eq.~(\ref{eq:collider_rdef}).
    }
    \label{tab:eight_squark_results}
    \begin{tabular}{ccccc}
        \hline\hline
        $m_{\widetilde H}$ [TeV] & $m_{\tilde q}$ [TeV] &
        $p_j^*$ [GeV] & $r_{\rm exp}$ & $r_{\rm obs}$ \\
        \hline
        1.10 & 1.15   & 48.9  & 0.46 & 0.44 \\
        1.10 & 1.20   & 95.8  & 0.39 & 0.42 \\
        1.10 & 1.50   & 346.7 & 0.10 & 0.11 \\
        1.10 & 1.80   & 563.9 & 0.08 & 0.06 \\
        \hline
        0.95 & 1.00   & 48.8  & 0.66 & 0.58 \\
        0.95 & 1.035  & 81.5  & 0.60 & 0.54 \\
        0.95 & 1.35   & 340.7 & 0.19 & 0.20 \\
        0.95 & 1.65   & 551.5 & 0.18 & 0.14 \\
        \hline\hline
    \end{tabular}
\end{table}

Table~\ref{tab:eight_squark_results} illustrates the competition between
decay kinematics and production rate. Larger mass gaps give harder decay
jets, while heavier squarks suffer from rapidly decreasing production cross
sections. None of the Run-2 points reaches exclusion, but the largest
sensitivity occurs when the squarks remain close in mass to the Higgsino and
the production rate is still sizable. Only this near-degenerate region can
therefore approach the HL-LHC exclusion sensitivity. We obtain a conditional
HL-LHC estimate by rescaling the Run-2 signal with the ratio of the squark-pair
production cross sections at $14$ and $13$~TeV. The acceptance and background
rate per unit luminosity are kept at their Run-2 values, as is the fractional
background uncertainty.

\begin{figure}[!htb]
    \centering
    \includegraphics[width=\textwidth]
    {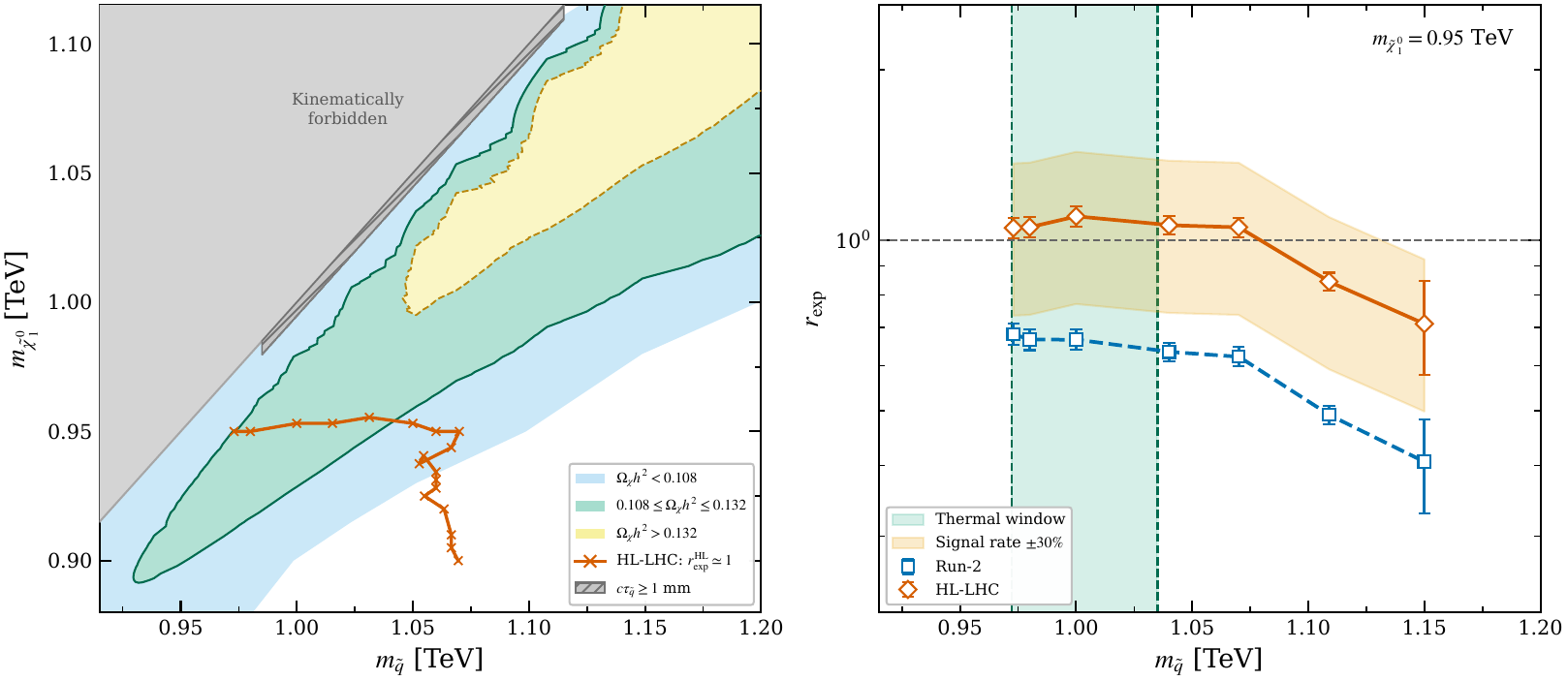}
    \caption{Relic density and conditional HL-LHC sensitivity for
light-flavor squarks. In the left panel, blue, green, and yellow show
$\Omega_\chi h^2$ below, within, and above $0.108$--$0.132$.
The relic-density band closes at low masses.
Gray hatching marks $c\tau_{\tilde q}\geq1$~mm. Crosses mark
simulated collider points, and the orange line joins those near
$r_{\rm exp}^{\rm HL}=1$. The right panel shows the $0.95$~TeV Higgsino
slice. Its pale-yellow band varies the signal rate by $\pm30\%$.
}
\label{fig:squark_hllhc_slice}
\end{figure}

Note that the same near-degeneracy changes the thermal history of the dark matter.
When the squark--Higgsino mass gap is small, the Boltzmann suppression of the
squark population during freeze-out is weak. Squark annihilation through the
strong interaction and squark--Higgsino coannihilation then increase the
effective annihilation rate and reduce the relic abundance relative to the
Higgsino-only result. As the mass gap grows, the squarks decouple from
freeze-out and the pure-Higgsino limit is recovered. We calculate the relic
abundance across the squark--Higgsino mass plane with the gluino and other
heavy scalars fixed at $5$~TeV. Sommerfeld enhancement and ground-state
bound-state formation are included in the no-transition approximation
\cite{Becker:2025vgq,Becker:2026icc}. Figure~\ref{fig:squark_hllhc_slice}
combines this calculation with the conditional HL-LHC sensitivity and
identifies the region that satisfies the relic-density requirement while
remaining accessible at the HL-LHC.

In the left panel, the green band marks the region with the observed relic
abundance. It is bounded by excessive depletion near the diagonal and by the
Higgsino-only result as the squarks decouple at larger mass splittings. The
band closes near
$(m_{\tilde q},m_{\tilde\chi_1^0})\simeq(0.93,0.89)$~TeV.
The narrow hatched strip next to the diagonal has
$c\tau_{\tilde q}\geq1$~mm at mass gaps of about $5.3$--$5.6$~GeV.
Away from this strip, the orange line meets the allowed thermal band
near $m_{\tilde\chi_1^0}=0.95$~TeV.
The variation of the sensitivity curve around $m_{\widetilde H}=0.925$~TeV is related to changes in the MB-C bin composition, where more signal events enter the high-significance five-jet categories.

The right panel illustrates this overlap for the
$m_{\chi_1^0}=0.95$~TeV slice. The relic-density requirement selects
$m_{\tilde q}=0.972$--$1.035$~TeV. The Run-2 MB-C ratios stay below
unity at the simulated points in this window. At the low-$m_{\tilde q}$ end,
$m_{\tilde q}=0.973$ and $0.980$~TeV both give
$r_{\rm exp}^{\rm HL}\simeq1.1$ in the conditional projection.
This part of the coannihilation region therefore reaches expected
$95\%$ CL exclusion sensitivity under the assumptions used here.


\subsubsection{Stop}
\label{subsec:rh_stop}

A stop provides a second QCD-produced channel, but unlike a light-flavor
squark its visible final state changes when the top channel opens. We use
the simple benchmark $\tilde t_1=\tilde t_R$, with the other colored
sparticles at $5$~TeV. Left--right mixing would change the relative decay
rates, while the right-handed limit isolates the top-Yukawa coupling to the
three two-body modes
\begin{equation}
    \tilde t_1\to
    t\,\tilde\chi_1^0,\qquad
    t\,\tilde\chi_2^0,\qquad
    b\,\tilde\chi_1^+ .
    \label{eq:stop_decays}
\end{equation}
For the main Higgsino benchmark, the neutral channels open at
\begin{equation}
    m_{\tilde t_1}
    \simeq m_{\tilde\chi_1^0}+m_t
    \simeq1.27~{\rm TeV}.
    \label{eq:stop_threshold}
\end{equation}
Below this threshold the stop decays almost entirely to
$b\tilde\chi_1^+$. Above it, the two neutral modes rapidly become
important, and stop pairs populate $t\bar t$, $tb$, and $b\bar b$
final states with missing transverse momentum. If
$B_t={\rm BR}(t\tilde\chi^0_1)+{\rm BR}(t\tilde\chi^0_2)$, the three
pair fractions are $B_t^2$, $2B_t(1-B_t)$, and $(1-B_t)^2$.

Since all three stop-pair final states contain $b$ jets,we therefore use the
CMS $b$-rich aggregate signal region A07~\cite{CMS:2019zmd} to follow
their contributions across the top threshold with a fixed selection.
A07 requires at least two jets and two $b$-tagged jets,
$H_T\geq600~{\rm GeV}$, and jet-based missing transverse momentum
$H_T^{\rm miss}\geq600~{\rm GeV}$. Its expected $95\%$ CL upper limit is
43 signal events.  Appendix~\ref{app:squark_stop_validation}
compares our event yields with public $T2tt$
($\tilde t\to t\tilde\chi_1^0$) and $T2bb$
($\tilde b\to b\tilde\chi_1^0$) benchmarks.

\begin{figure}[!htb]
    \centering
    \includegraphics[width=0.90\textwidth]{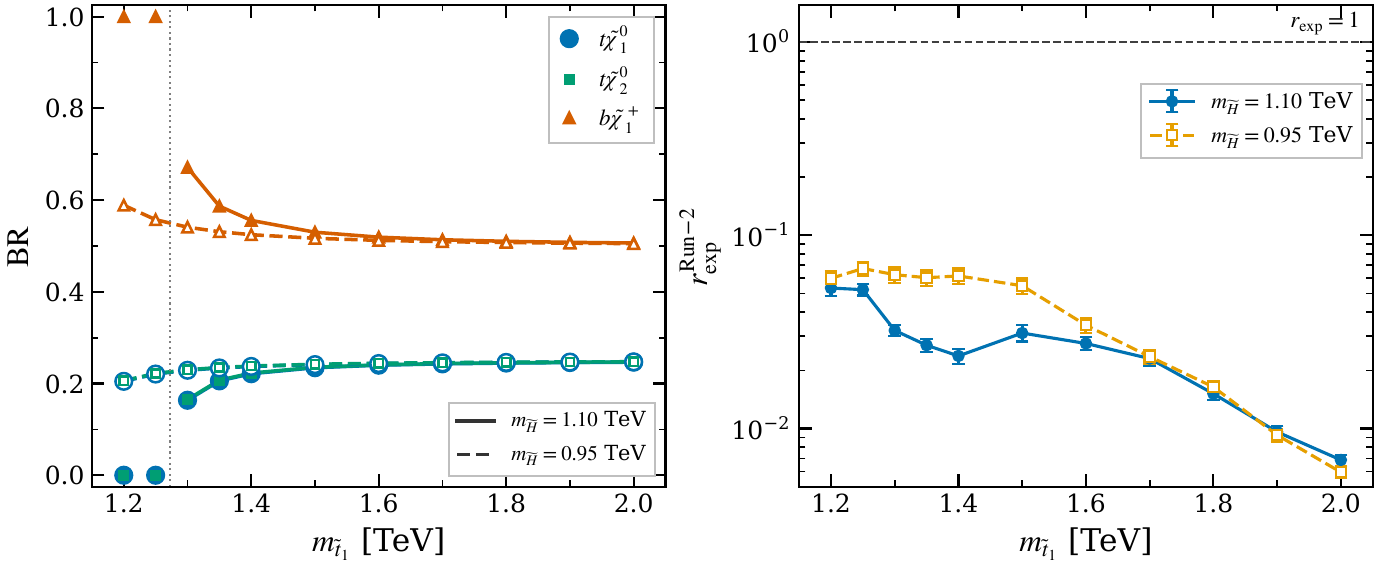}
    \caption{ One dimensional plot for right-handed stop.
    The left panel shows the branching fractions  for
    $m_{\widetilde H}=0.95$ and $1.10$~TeV. The right panel shows the
    expected Run-2 CMS sensitivity. The dotted line marks the opening
    of the $t\tilde\chi_{1,2}^0$ modes for the $1.10$~TeV benchmark. Error
    bars show the signal Monte Carlo uncertainty.
    }
    \label{fig:stop_decay_a07}
\end{figure}

Figure~\ref{fig:stop_decay_a07} shows the branching fractions and the
corresponding sensitivity for the two Higgsino benchmarks. The left
panel displays how the top threshold changes the stop decay pattern, while
the right panel shows the resulting expected sensitivity.
For $m_{\widetilde H}=1.10$~TeV, the top--neutralino channels open in
the scan. At $m_{\tilde t_1}=1.3~{\rm TeV}$, \textsc{SDECAY} gives
${\rm BR}(t\tilde\chi_1^0,t\tilde\chi_2^0,b\tilde\chi_1^+)
=(0.164,0.165,0.671)$. As the phase-space suppression decreases, these
fractions approach the Higgsino coupling ratio $1:1:2$. The stop-pair final state therefore changes from an almost pure $bb$ topology below threshold to a mixture of $tt$, $tb$, and $bb$ above it.

The right panel shows the corresponding Run-2 sensitivity. The new neutral
modes initially have little phase space, so the acceptance decreases before
the harder decay products raise it again. Together with the falling
production cross section, this delayed efficiency turn-on produces the
minimum near $1.4$~TeV and the modest recovery at $1.5$~TeV. For
$m_{\widetilde H}=0.95$~TeV, the threshold lies near $1.12$~TeV, below the
plotted range, and the curve has no analogous feature. At larger stop
masses, the production rate becomes the dominant limitation. Overall, the stop sensitivity remains limited by the interplay between
threshold kinematics and the falling production rate. For the two Higgsino
benchmarks considered, $r_{\rm exp}<0.07$ throughout the studied stop-mass
range, far below the sensitivity required for exclusion by the current Run-2
search. Even an optimistic statistics-dominated extrapolation to $14$~TeV
with $3~{\rm ab}^{-1}$, including the increase in the stop-pair production
cross section, gives $r_{\rm exp}\lesssim0.4$. The stop channel therefore
remains beyond the expected exclusion reach of this search at the HL-LHC.

\subsubsection{Sbottom}
\label{subsec:sbottom}

A right-handed sbottom provides another third-generation colored portal to
the Higgsino sector. Its decay pattern changes when the charged mode
$\tilde b_R\to t\tilde\chi_1^-$ opens. We take
$\tilde b_1=\tilde b_R$ with the other colored sparticles fixed at
$5$~TeV. For $m_{\widetilde H}=1.10$~TeV, the neutral modes
$\tilde b_R\to b\tilde\chi^0_{1,2}$ are open throughout the mass range
studied, while the top mode opens near
$m_{\tilde b_R}=1.27$~TeV. Above this threshold, sbottom pairs produce
$tt$, $tb$, and $bb$ final states with missing transverse momentum. The
size of the top-rich component is controlled by the branching fraction of
the charged mode.

In the pure-Higgsino limit, the charged and summed-neutral widths carry the
same factor of $y_b^2$. Their relative size is therefore mainly determined
by the available phase space,
\begin{equation}
B_t^R\equiv {\rm BR}(\tilde b_R\to t\tilde\chi_1^-)
\simeq \frac{F_t}{F_t+F_0},
\end{equation}
where
\begin{equation}
F_f=\lambda^{1/2}(1,r_{\widetilde H},r_f)
(1-r_{\widetilde H}-r_f),\qquad
r_i=\frac{m_i^2}{m_{\tilde b_R}^2}.
\end{equation}
Here
$\lambda(1,x,y)=(1-x-y)^2-4xy$, while $F_t$ and $F_0$ denote
$F_f$ evaluated with $m_f=m_t$ and $m_f=0$ respectively. At
$m_{\tilde b_R}=1.25$~TeV the top mode is closed, and the two neutral modes
have nearly equal branching fractions. SDECAY gives
$B_t^R=0.326$ and $0.469$ at $1.30$ and $1.50$~TeV respectively. As the
top-mass suppression decreases, the charged and summed-neutral modes
approach comparable branching fractions.

To capture the change in decay topology, we choose representative masses
below and above the top threshold and apply the same $b$-rich search strategy
used for the stop benchmark. The resulting sensitivities are
$r_{\rm exp}=0.05$, $0.03$, and $0.03$ at
$m_{\tilde b_R}=1.25$, $1.30$, and $1.50$~TeV respectively. These values
are close to those of the right-handed stop at the corresponding masses and
remain well below unity. Although the opening of the charged Higgsino mode
changes the sbottom decay topology above the top threshold, it does not
compensate for the compressed kinematics and the limited production rate.
The largest value, $r_{\rm exp}=0.05$, is far below the sensitivity required
for exclusion by the current Run-2 search. An optimistic statistics-dominated
extrapolation to $14$~TeV with $3~{\rm ab}^{-1}$ gives
$r_{\rm exp}\lesssim0.3$, leaving the right-handed sbottom beyond the expected
exclusion reach of this search at the HL-LHC.

\subsection{Sleptons}
\label{subsec:sleptons}

Slepton pairs provide a clean dilepton route to the heavy Higgsino, but
their electroweak production rate is much smaller than the QCD rates. Drell--Yan production is set by gauge couplings, whereas the decay
to a pure Higgsino is proportional to the lepton Yukawa coupling. The
small Yukawa coupling makes the width narrow but does not reduce the
branching fraction when no competing channel is open. We restrict the
discussion to unmixed selectrons and smuons, whose decays are prompt.
The charged-slepton branching fractions are
\begin{equation}
\begin{split}
    {\rm BR}(\tilde\ell_L\to\ell\tilde\chi^0_1,
             \ell\tilde\chi^0_2)&\simeq(0.50,0.50),\\
    {\rm BR}(\tilde\ell_R\to\ell\tilde\chi^0_1,
             \ell\tilde\chi^0_2,\nu\tilde\chi^-_1)
             &\simeq(0.25,0.25,0.50).
\end{split}
    \label{eq:slepton_branching}
\end{equation}
so the dilepton fraction is one for a left-handed slepton pair and $1/4$
for a right-handed pair, before kinematic selection.

A near-threshold benchmark illustrates the limited event yield of this channel. 
We take
$m_{\tilde\ell}=1.2$~TeV, which is close to the kinematic threshold while 
retaining a relatively large production rate. At $14$~TeV, the estimated pair-production cross sections per 
flavor are about
$4.6\times10^{-3}$~fb for $\tilde\ell_L$ and
$2.0\times10^{-3}$~fb for $\tilde\ell_R$.
For $3~{\rm ab}^{-1}$, this corresponds to about 14 and 6 produced slepton 
pairs respectively. Including four degenerate states
$\tilde e_{L,R}$ and $\tilde\mu_{L,R}$, Eq.~(\ref{eq:slepton_branching}) 
gives approximately 30 dilepton events before detector acceptance and analysis 
selections. The surviving signal yield would therefore be substantially 
smaller, making this channel challenging for a robust exclusion.

The limited sensitivity is further aggravated by the compressed kinematics. 
For
$m_{\tilde\ell}=1.2$~TeV and
$m_{\tilde\chi}\simeq1.1$~TeV, the lepton momentum in the slepton rest frame is $95.8~{\rm GeV}$. Increasing the slepton mass improves the lepton hardness but rapidly suppresses the electroweak production rate. Therefore, unlike colored superpartners, sleptons do not provide a high-rate portal to TeV-scale Higgsinos at hadron colliders. Even at the HL-LHC, the combination of small production cross sections and compressed decay kinematics limits the discovery potential of this  channel. Although sleptons offer a clean dilepton signature, they do not reach expected Run-2 exclusion in the end.

\subsection{Heavier electroweakinos}
\label{subsec:heavier_electroweakinos}

The splitting between two neutral Higgsino-like states can arise from very heavy gauginos, as assumed in the colored-parent benchmarks above, or from a cancellation between opposite-sign bino and wino contributions. In the
latter case, the direct-detection scan gives $|M_1|\gtrsim1.68$~TeV and
$M_2\gtrsim2.30$~TeV, with the two bounds reached at different points.
Direct production is limited by these high mass scales. At a wino mass
of $2$~TeV, the 14~TeV pure-wino associated-production cross section is
$9.9\times10^{-3}$~fb~\cite{LHCSUSYXS:Wino14}, giving about $30$ pairs
before cuts at $3~{\rm ab}^{-1}$. The allowed wino mass scale is higher,
and a nearly pure bino has no comparable unsuppressed Drell--Yan
pair-production channel when squarks are heavy. Therefore, direct production of heavier electroweakinos remains a challenging probe of the heavy Higgsino.

\FloatBarrier

\section{Conclusions}
\label{sec:conclusions}

We have studied LHC tests of the LZ-motivated Higgsino dark
matter scenario. The relic abundance fixes the Higgsino mass near $1.1$~TeV,
and the LZ recoil selects a neutral splitting of a few hundred keV. Current
direct searches do not reach this heavy and compressed multiplet. The
splitting constrains the bino and wino sectors but permits TeV-scale colored
superpartners and sleptons, motivating searches for their cascade decays into
Higgsinos.
Gluino pair production provides the strongest sensitivity.
For a $1.1$~TeV Higgsino, the CMS Run-2 search excludes gluinos below
$2.21$~TeV, while the HL-LHC exclusion and discovery reaches are
$2.45$--$2.60$~TeV and $2.26$--$2.37$~TeV respectively. The light-flavor
squark coannihilation region lies beyond the Run-2 reach, and only its
lower-mass edge approaches HL-LHC exclusion under optimistic assumptions.
The stop and sbottom channels remain far below the current exclusion threshold
and do not reach expected HL-LHC exclusion. Slepton and heavier-electroweakino
channels are also inaccessible at the HL-LHC because of the small rates and
compressed decay kinematics. Overall, the production and decay of heavier
superpartners may yield observable signals at the current LHC and the HL-LHC,
providing a complementary test of the LZ-motivated Higgsino interpretation.

\addcontentsline{toc}{section}{Acknowledgments}
\acknowledgments
This work was supported by the National Natural Science Foundation of China (Grant No. 12335005), by the Natural Science Foundation of Henan Province (Grant No. 262300421233), and by the PI Research Fund of Henan Normal University (Grant No. 5101029470335).

\appendix
\section{Recast validation}
\label{app:recast_validation}

\subsection{Gluino recast validation}
\label{app:gluino_validation}

\begin{figure}[htbp]
    \centering
    \includegraphics[width=\textwidth]{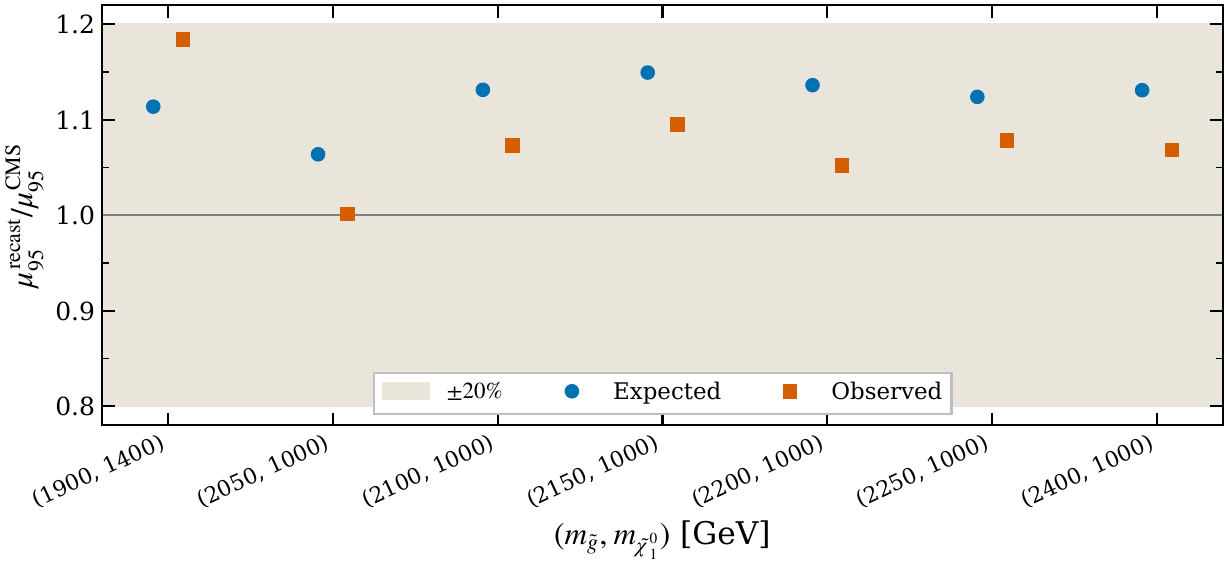}
    \caption{ The ratios of signal strength between recast and CMS results for seven $T1tttt$ benchmarks. The shaded strip marks the $\pm20\%$ validation range.
    }
    \label{fig:gluino_validation}
\end{figure}

In this appendix, we show in Figure~\ref{fig:gluino_validation} seven public $T1tttt$
benchmarks used to validate the CMS prompt-$M_{T2}$\cite{CMS:2019ybf}.
The $(1900,1400)$~GeV benchmark uses the low-$H_T$ covariance block, while
the six $m_{\tilde\chi^0_1}=1000$~GeV points use the high-$H_T$ block.
The largest deviations in the expected and observed upper limits are
$14.95\%$ and $18.40\%$, respectively.

On the $1000$~GeV daughter slice, the expected and observed mass
crossings are $2083.0$ and $2183.3$~GeV, compared with the CMS reference
values $2116.0$ and $2201.3$~GeV. The corresponding differences are
$1.56\%$ and $0.82\%$.

\FloatBarrier

\subsection{Squark and stop benchmarks}
\label{app:squark_stop_validation}

To check the MB-C recast of the ATLAS jets plus missing-energy search~\cite{ATLAS:2020syg}, we use the public $T2qq$ benchmark with degenerate first- and second-generation light-flavor squarks at $m_{\tilde q}=1.7$~TeV and $m_{\tilde\chi_1^0}=0.5$~TeV~\cite{Lara:2025cpm,Drees:2013wra,Dercks:2016npn}. Although this spectrum is not compressed, published CheckMATE yields are available for all 18 MB-C bins, allowing a direct check of our event selection. Our yields agree within $20\%$ in 16 of these bins. In the four bins most relevant to the expected limit, they agree within $6\%$.

We also compare the expected limits using the ATLAS background-only likelihood and signal patch from HEPData.\footnote{\url{https://doi.org/10.17182/hepdata.95664}.} The official signal patch gives a median expected $95\%$ CL upper limit of $\mu_{95}=1.673$. Replacing its signal yields with our simulated ones gives $1.460$. The relative difference is around $12.7\%$.

For the stop analysis, we check the number of events passing the CMS A07 selection against public $T2tt$ and $T2bb$ benchmarks~\cite{CMS:2019zmd}. Our yields agree with the reference values at the $10$--$15\%$ level. 

\FloatBarrier

\bibliographystyle{CitationStyle}
\bibliography{myrefs}

@article{Nagata:2014wma,
    author = "Nagata, Natsumi and Shirai, Satoshi",
    title = "{Higgsino Dark Matter in High-Scale Supersymmetry}",
    eprint = "1410.4549",
    archivePrefix = "arXiv",
    primaryClass = "hep-ph",
    reportNumber = "DESY-14-180, FTPI-MINN-14-37, IPMU14-0320",
    doi = "10.1007/JHEP01(2015)029",
    journal = "JHEP",
    volume = "01",
    pages = "029",
    year = "2015"
}

@article{LZ:2026axp,
    author = "Akerib, D. S. and others",
    collaboration = "LZ",
    title = "{Search for dark matter particle interactions in an extended nuclear recoil energy window with the LUX-ZEPLIN (LZ) experiment}",
    eprint = "2609.02823",
    archivePrefix = "arXiv",
    primaryClass = "hep-ex",
    doi = "10.17182/hepdata.182472.v1",
    month = "9",
    year = "2026"
}

@article{Fan:2026kxx,
    author = "Fan, JiJi and Reece, Matthew",
    title = "{Higgsino Above the Sea of Fog}",
    eprint = "2609.01504",
    archivePrefix = "arXiv",
    primaryClass = "hep-ph",
    month = "9",
    year = "2026"
}

@article{Freese:2026sga,
    author = "Freese, Katherine and Theodosopoulos, Dionysios P.",
    title = "{Higgsino Dark Matter Interpretation of the LUX-ZEPLIN 248 keV Nuclear-Recoil Event}",
    eprint = "2609.01583",
    archivePrefix = "arXiv",
    primaryClass = "hep-ph",
    month = "9",
    year = "2026"
}

@article{Wu:2026nhi,
    author = "Wu, Lei and Zhang, Yang and Zhu, Bin",
    title = "{TeV Higgsino Dark Matter from LZ Nuclear Recoil to Fermi-LAT Gamma Rays}",
    eprint = "2609.01590",
    archivePrefix = "arXiv",
    primaryClass = "hep-ph",
    month = "9",
    year = "2026"
}

@article{Pospelov:2026ewn,
    author = "Pospelov, Maxim and Ramani, Harikrishnan",
    title = "{Strong Constraints on Higgsino Dark Matter from Solar Capture}",
    eprint = "2609.02775",
    archivePrefix = "arXiv",
    primaryClass = "hep-ph",
    month = "9",
    year = "2026"
}

@article{Rodd:2026tyn,
    author = "Rodd, Nicholas L. and Safdi, Benjamin R. and Slatyer, Tracy R. and Xu, Weishuang Linda",
    title = "{Confronting the Higgsino Interpretation of the LZ Event with the High-Energy Sideband}",
    eprint = "2609.04175",
    archivePrefix = "arXiv",
    primaryClass = "hep-ph",
    month = "9",
    year = "2026"
}

@article{Low:2014cba,
    author = "Low, Matthew and Wang, Lian-Tao",
    title = "{Neutralino dark matter at 14 TeV and 100 TeV}",
    eprint = "1404.0682",
    archivePrefix = "arXiv",
    primaryClass = "hep-ph",
    reportNumber = "EFI-14-6",
    doi = "10.1007/JHEP08(2014)161",
    journal = "JHEP",
    volume = "08",
    pages = "161",
    year = "2014"
}

@article{Han:2018wus,
    author = "Han, Tao and Mukhopadhyay, Satyanarayan and Wang, Xing",
    title = "{Electroweak Dark Matter at Future Hadron Colliders}",
    eprint = "1805.00015",
    archivePrefix = "arXiv",
    primaryClass = "hep-ph",
    reportNumber = "PITT-PACC-1806",
    doi = "10.1103/PhysRevD.98.035026",
    journal = "Phys. Rev. D",
    volume = "98",
    number = "3",
    pages = "035026",
    year = "2018"
}

@article{Capdevilla:2024bwt,
    author = "Capdevilla, Rodolfo and Meloni, Federico and Zurita, Jose",
    title = "{Discovering Electroweak Interacting Dark Matter at Muon Colliders Using Soft Tracks}",
    eprint = "2405.08858",
    archivePrefix = "arXiv",
    primaryClass = "hep-ph",
    reportNumber = "FERMILAB-PUB-23-0832-T, DESY-24-069",
    doi = "10.1103/PhysRevLett.134.181802",
    journal = "Phys. Rev. Lett.",
    volume = "134",
    number = "18",
    pages = "181802",
    year = "2025"
}

@article{Bhattiprolu:2025zwt,
    author = "Bhattiprolu, Prudhvi N. and Martin, Stephen P. and Wells, James D.",
    title = "{Chasing Higgsino dark matter at colliders in the neutrino fog era}",
    eprint = "2512.12457",
    archivePrefix = "arXiv",
    primaryClass = "hep-ph",
    doi = "10.1103/m7qf-j8j6",
    journal = "Phys. Rev. D",
    volume = "113",
    number = "9",
    pages = "095029",
    year = "2026"
}

@article{Cheung:2026byg,
    author = "Cheung, Kingman and Kang, Sin Kyu and Kumar, Ranjeet",
    title = "{From LUX-ZEPLIN to Colliders: Probing Higgsino Dark Matter}",
    eprint = "2609.08712",
    archivePrefix = "arXiv",
    primaryClass = "hep-ph",
    month = "9",
    year = "2026"
}

@article{Alwall:2014hca,
    author = "Alwall, J. and Frederix, R. and Frixione, S. and Hirschi, V. and Maltoni, F. and Mattelaer, O. and Shao, H. -S. and Stelzer, T. and Torrielli, P. and Zaro, M.",
    title = "{The automated computation of tree-level and next-to-leading order differential cross sections, and their matching to parton shower simulations}",
    eprint = "1405.0301",
    archivePrefix = "arXiv",
    primaryClass = "hep-ph",
    reportNumber = "CERN-PH-TH-2014-064, CP3-14-18, LPN14-066, MCNET-14-09, ZU-TH-14-14",
    doi = "10.1007/JHEP07(2014)079",
    journal = "JHEP",
    volume = "07",
    pages = "079",
    year = "2014"
}

@article{Sjostrand:2014zea,
    author = {Sj{\"o}strand, Torbj{\"o}rn and Ask, Stefan and Christiansen, Jesper R. and Corke, Richard and Desai, Nishita and Ilten, Philip and Mrenna, Stephen and Prestel, Stefan and Rasmussen, Christine O. and Skands, Peter Z.},
    title = "{An introduction to PYTHIA 8.2}",
    eprint = "1410.3012",
    archivePrefix = "arXiv",
    primaryClass = "hep-ph",
    reportNumber = "LU-TP-14-36, MCNET-14-22, CERN-PH-TH-2014-190, FERMILAB-PUB-14-316-CD, DESY-14-178, SLAC-PUB-16122",
    doi = "10.1016/j.cpc.2015.01.024",
    journal = "Comput. Phys. Commun.",
    volume = "191",
    pages = "159--177",
    year = "2015"
}

@article{Drees:2013wra,
    author = "Drees, Manuel and Dreiner, Herbi and Schmeier, Daniel and Tattersall, Jamie and Kim, Jong Soo",
    title = "{CheckMATE: Confronting your Favourite New Physics Model with LHC Data}",
    eprint = "1312.2591",
    archivePrefix = "arXiv",
    primaryClass = "hep-ph",
    reportNumber = "ADP-13-29-T849, IFT-UAM-CSIC-13-133, ADP-13-29-T849-IFT-UAM-CSIC-13-133",
    doi = "10.1016/j.cpc.2014.10.018",
    journal = "Comput. Phys. Commun.",
    volume = "187",
    pages = "227--265",
    year = "2015"
}

@article{Dercks:2016npn,
    author = "Dercks, Daniel and Desai, Nishita and Kim, Jong Soo and Rolbiecki, Krzysztof and Tattersall, Jamie and Weber, Torsten",
    title = "{CheckMATE 2: From the model to the limit}",
    eprint = "1611.09856",
    archivePrefix = "arXiv",
    primaryClass = "hep-ph",
    reportNumber = "CTPU-16-36, CSIC-16-116, TTK-16-47",
    doi = "10.1016/j.cpc.2017.08.021",
    journal = "Comput. Phys. Commun.",
    volume = "221",
    pages = "383--418",
    year = "2017"
}

@article{CMS:2019ybf,
    author = "Sirunyan, Albert M and others",
    collaboration = "CMS",
    title = "{Searches for physics beyond the standard model with the $M_\mathrm{T2}$ variable in hadronic final states with and without disappearing tracks in proton-proton collisions at $\sqrt{s}=$ 13 TeV}",
    eprint = "1909.03460",
    archivePrefix = "arXiv",
    primaryClass = "hep-ex",
    reportNumber = "CMS-SUS-19-005, CERN-EP-2019-180",
    doi = "10.1140/epjc/s10052-019-7493-x",
    journal = "Eur. Phys. J. C",
    volume = "80",
    number = "1",
    pages = "3",
    year = "2020"
}

@article{ATLAS:2020syg,
    author = "Aad, Georges and others",
    collaboration = "ATLAS",
    title = "{Search for squarks and gluinos in final states with jets and missing transverse momentum using 139 fb$^{-1}$ of $\sqrt{s}$ =13 TeV $pp$ collision data with the ATLAS detector}",
    eprint = "2010.14293",
    archivePrefix = "arXiv",
    primaryClass = "hep-ex",
    reportNumber = "CERN-EP-2020-166",
    doi = "10.1007/JHEP02(2021)143",
    journal = "JHEP",
    volume = "02",
    pages = "143",
    year = "2021"
}

@article{deFavereau:2013fsa,
    author = "de Favereau, J. and Delaere, C. and Demin, P. and Giammanco, A. and Lema{\^\i}tre, V. and Mertens, A. and Selvaggi, M.",
    collaboration = "DELPHES 3",
    title = "{DELPHES 3, A modular framework for fast simulation of a generic collider experiment}",
    eprint = "1307.6346",
    archivePrefix = "arXiv",
    primaryClass = "hep-ex",
    doi = "10.1007/JHEP02(2014)057",
    journal = "JHEP",
    volume = "02",
    pages = "057",
    year = "2014"
}

@article{Bierlich:2022pfr,
    author = "Bierlich, Christian and others",
    title = "{A comprehensive guide to the physics and usage of PYTHIA 8.3}",
    eprint = "2203.11601",
    archivePrefix = "arXiv",
    primaryClass = "hep-ph",
    reportNumber = "LU-TP 22-16, MCNET-22-04, FERMILAB-PUB-22-227-SCD",
    doi = "10.21468/SciPostPhysCodeb.8",
    journal = "SciPost Phys. Codeb.",
    volume = "2022",
    pages = "8",
    year = "2022"
}

@article{Muhlleitner:2003vg,
    author = "Muhlleitner, M. and Djouadi, A. and Mambrini, Y.",
    title = "{SDECAY: A Fortran code for the decays of the supersymmetric particles in the MSSM}",
    eprint = "hep-ph/0311167",
    archivePrefix = "arXiv",
    reportNumber = "CERN-TH-2003-252, PM-03-22, PSI-PR-03-17",
    doi = "10.1016/j.cpc.2005.01.012",
    journal = "Comput. Phys. Commun.",
    volume = "168",
    pages = "46--70",
    year = "2005"
}

@article{Alloul:2013bka,
    author = "Alloul, Adam and Christensen, Neil D. and Degrande, C{\'e}line and Duhr, Claude and Fuks, Benjamin",
    title = "{FeynRules  2.0 - A complete toolbox for tree-level phenomenology}",
    eprint = "1310.1921",
    archivePrefix = "arXiv",
    primaryClass = "hep-ph",
    reportNumber = "CERN-PH-TH-2013-239, MCNET-13-14, IPPP-13-71, DCPT-13-142, PITT-PACC-1308",
    doi = "10.1016/j.cpc.2014.04.012",
    journal = "Comput. Phys. Commun.",
    volume = "185",
    pages = "2250--2300",
    year = "2014"
}

@article{Degrande:2011ua,
    author = "Degrande, Celine and Duhr, Claude and Fuks, Benjamin and Grellscheid, David and Mattelaer, Olivier and Reiter, Thomas",
    title = "{UFO - The Universal FeynRules Output}",
    eprint = "1108.2040",
    archivePrefix = "arXiv",
    primaryClass = "hep-ph",
    reportNumber = "CP3-11-25, IPHC-PHENO-11-04, IPPP-11-39, DCPT-11-78, MPP-2011-68",
    doi = "10.1016/j.cpc.2012.01.022",
    journal = "Comput. Phys. Commun.",
    volume = "183",
    pages = "1201--1214",
    year = "2012"
}

@article{CMS:2019zmd,
    author = "Collaboration, The Cms and others",
    collaboration = "CMS",
    title = "{Search for supersymmetry in proton-proton collisions at 13 TeV in final states with jets and missing transverse momentum}",
    eprint = "1908.04722",
    archivePrefix = "arXiv",
    primaryClass = "hep-ex",
    reportNumber = "CMS-SUS-19-006, CERN-EP-2019-152",
    doi = "10.1007/JHEP10(2019)244",
    journal = "JHEP",
    volume = "10",
    pages = "244",
    year = "2019"
}

@article{Araz:2023bwx,
    author = "Araz, Jack Y.",
    title = "{Spey: Smooth inference for reinterpretation studies}",
    eprint = "2307.06996",
    archivePrefix = "arXiv",
    primaryClass = "hep-ph",
    reportNumber = "IPPP/23/34",
    doi = "10.21468/SciPostPhys.16.1.032",
    journal = "SciPost Phys.",
    volume = "16",
    number = "1",
    pages = "032",
    year = "2024"
}

@article{Heinrich:2021gyp,
    author = "Heinrich, Lukas and Feickert, Matthew and Stark, Giordon and Cranmer, Kyle",
    title = "{pyhf: pure-Python implementation of HistFactory statistical models}",
    doi = "10.21105/joss.02823",
    journal = "J. Open Source Softw.",
    volume = "6",
    number = "58",
    pages = "2823",
    year = "2021"
}

@article{Cranmer:2012sba,
    author = "Cranmer, Kyle and Lewis, George and Moneta, Lorenzo and Shibata, Akira and Verkerke, Wouter",
    collaboration = "ROOT",
    title = "{HistFactory: A tool for creating statistical models for use with RooFit and RooStats}",
    reportNumber = "CERN-OPEN-2012-016",
    month = "6",
    year = "2012"
}

@article{Cowan:2010js,
    author = "Cowan, Glen and Cranmer, Kyle and Gross, Eilam and Vitells, Ofer",
    title = "{Asymptotic formulae for likelihood-based tests of new physics}",
    eprint = "1007.1727",
    archivePrefix = "arXiv",
    primaryClass = "physics.data-an",
    doi = "10.1140/epjc/s10052-011-1554-0",
    journal = "Eur. Phys. J. C",
    volume = "71",
    pages = "1554",
    year = "2011",
    note = "[Erratum: Eur.Phys.J.C 73, 2501 (2013)]"
}

@article{Read:2002hq,
    author = "Read, Alexander L.",
    editor = "Whalley, M. R. and Lyons, L.",
    title = "{Presentation of search results: The $CL_s$ technique}",
    doi = "10.1088/0954-3899/28/10/313",
    journal = "J. Phys. G",
    volume = "28",
    pages = "2693--2704",
    year = "2002"
}

@article{Lara:2025cpm,
    author = "Lara, I{\~n}aki and Rolbiecki, Krzysztof",
    title = "{Implementation of full and simplified likelihoods in CheckMATE}",
    eprint = "2507.08565",
    archivePrefix = "arXiv",
    primaryClass = "hep-ph",
    doi = "10.1140/epjc/s10052-026-15414-8",
    journal = "Eur. Phys. J. C",
    volume = "86",
    number = "3",
    pages = "221",
    year = "2026"
}

@article{Su:2026rwz,
    author = "Su, Liangliang and Yang, Jin Min and Yang, Wen-Na",
    title = "{Inelastic Dark Matter Signature at High Recoil Energy in LUX-ZEPLIN and CRESST}",
    eprint = "2609.01475",
    archivePrefix = "arXiv",
    primaryClass = "hep-ph",
    month = "9",
    year = "2026"
}

@article{Lou:2026idn,
    author = "Lou, Yuanchao and Lu, Chih-Ting",
    title = "{Fermionic Dark Matter Absorption and the High-Energy Event in LUX-ZEPLIN}",
    eprint = "2609.01592",
    archivePrefix = "arXiv",
    primaryClass = "hep-ph",
    month = "9",
    year = "2026"
}

@article{Yin:2026jnn,
    author = "Yin, Wen",
    title = "{A PQ-Symmetric High-Scale SUSY Interpretation of the LZ High-Energy Recoil}",
    eprint = "2609.01892",
    archivePrefix = "arXiv",
    primaryClass = "hep-ph",
    month = "9",
    year = "2026"
}

@article{Nomura:2026qyq,
    author = "Nomura, Yasunori",
    title = "{Dark Matter as the Z{\_}2 Partner of the Standard Model Higgs Boson}",
    eprint = "2609.02505",
    archivePrefix = "arXiv",
    primaryClass = "hep-ph",
    reportNumber = "RIKEN-iTHEMS-Report-26",
    month = "9",
    year = "2026"
}

@article{DiMauro:2026ldr,
    author = "Di Mauro, Mattia",
    title = "{Dark Matter at the Kinematic Edge: Interpreting the 248 keV LZ Nuclear-Recoil Candidate}",
    eprint = "2609.02608",
    archivePrefix = "arXiv",
    primaryClass = "hep-ph",
    month = "9",
    year = "2026"
}

@article{Visinelli:2026kgt,
    author = "Visinelli, Luca",
    title = "{A Peccei-Quinn Origin for Inelastic Electroweak Dark Matter after LUX-ZEPLIN}",
    eprint = "2609.02807",
    archivePrefix = "arXiv",
    primaryClass = "hep-ph",
    month = "9",
    year = "2026"
}

@article{Yamashita:2026ump,
    author = "Yamashita, Kimiko",
    title = "{Inelastic Dark Photon Dark Matter for the LUX-ZEPLIN High-Recoil Event and the Galactic Halo Gamma-Ray Excess}",
    eprint = "2609.02868",
    archivePrefix = "arXiv",
    primaryClass = "hep-ph",
    month = "9",
    year = "2026"
}

@article{Chattopadhyay:2026ryw,
    author = "Chattopadhyay, Utpal and Das, Debottam and Puri, Rahul and Roy, Joydeep",
    title = "{Sub-TeV Singlino Dark Matter in light from Sagittarius A$^\ast$ and LUX-ZEPLIN Nuclear-Recoil Event}",
    eprint = "2609.02994",
    archivePrefix = "arXiv",
    primaryClass = "hep-ph",
    month = "9",
    year = "2026"
}

@article{Smirnov:2026aqk,
    author = "Smirnov, Juri and Griffith, Spencer and Beacom, John F.",
    title = "{Inelastic Signatures of Electroweak Dark Matter}",
    eprint = "2609.04144",
    archivePrefix = "arXiv",
    primaryClass = "hep-ph",
    month = "9",
    year = "2026"
}

@article{Du:2026guj,
    author = "Du, Xiaokang and Wang, Fei",
    title = "{TeV Higgsino Interpretation of the LZ High-Recoil Event with Intermediate-Scale Electroweak Gauginos}",
    eprint = "2609.04163",
    archivePrefix = "arXiv",
    primaryClass = "hep-ph",
    month = "9",
    year = "2026"
}

@article{McCabe:2026crm,
    author = "McCabe, Christopher",
    title = "{Seasonal dark matter from the LUX-ZEPLIN high-energy event}",
    eprint = "2609.04181",
    archivePrefix = "arXiv",
    primaryClass = "hep-ph",
    month = "9",
    year = "2026"
}

@article{Jeesun:2026vzo,
    author = "Jeesun, Sk and Majumdar, Anirban",
    title = "{Atmospheric neutrino up-scattering explanation of LZ 2026 excess}",
    eprint = "2609.04185",
    archivePrefix = "arXiv",
    primaryClass = "hep-ph",
    month = "9",
    year = "2026"
}

@article{Unwin:2026rdp,
    author = "Unwin, James",
    title = "{Axion Portal Dark Matter and the LUX-ZEPLIN High-Recoil Event}",
    eprint = "2609.04186",
    archivePrefix = "arXiv",
    primaryClass = "hep-ph",
    month = "9",
    year = "2026"
}

@article{Dent:2026bji,
    author = "Dent, James B. and Newstead, Jayden L.",
    title = "{Exothermic and Endothermic Inelastic Dark Matter Interpretations at LZ: Sideband Constraints and Future Prospects}",
    eprint = "2609.04673",
    archivePrefix = "arXiv",
    primaryClass = "hep-ph",
    month = "9",
    year = "2026"
}

@article{deLima:2026shq,
    author = "de Lima, Carlos Henrique",
    title = "{Exothermic Dark Matter at LZ}",
    eprint = "2609.05204",
    archivePrefix = "arXiv",
    primaryClass = "hep-ph",
    month = "9",
    year = "2026"
}

@article{Gu:2026vto,
    author = "Gu, Guanhua and Li, Lingfeng and Tang, Shao-Song and Xu, Yongheng",
    title = "{Inelastic from the Other Side: Xenon Excitation Signals in Light of the LZ High-Recoil Event}",
    eprint = "2609.05291",
    archivePrefix = "arXiv",
    primaryClass = "hep-ph",
    month = "9",
    year = "2026"
}

@article{Baer:2026fpy,
    author = "Baer, Howard and Barger, Vernon",
    title = "{Exothermic dark matter and the 248 keV nuclear recoil in LUX-ZEPLIN}",
    eprint = "2609.06153",
    archivePrefix = "arXiv",
    primaryClass = "hep-ph",
    month = "9",
    year = "2026"
}

@article{Lee:2026wof,
    author = "Lee, Hyun Min",
    title = "{Inelastic dark matter and baryon flavor symmetry in light of LUX-ZEPLIN (LZ) experiment}",
    eprint = "2609.06171",
    archivePrefix = "arXiv",
    primaryClass = "hep-ph",
    month = "9",
    year = "2026"
}

@article{Wang:2026ytg,
    author = "Wang, Lei and Xiao, Yang",
    title = "{The Inert Doublet Model of Dark Matter and the LUX-ZEPLIN High-Recoil Event}",
    eprint = "2609.06571",
    archivePrefix = "arXiv",
    primaryClass = "hep-ph",
    month = "9",
    year = "2026"
}

@article{Yang:2026wpb,
    author = "Yang, Meiwen and Wu, Quan-feng and Tsai, Yue-Lin Sming and Fan, Yi-Zhong",
    title = "{Multi-Messenger and Paleo-Detector Probes of the LZ Dark Matter Signal}",
    eprint = "2609.06640",
    archivePrefix = "arXiv",
    primaryClass = "hep-ph",
    month = "9",
    year = "2026"
}

@article{Liang:2026coz,
    author = "Liang, Jin-Han and Liu, Zuowei and Tran, Van Que and Xu, Yongheng",
    title = "{LZ Nuclear-Recoil Excess from Boosted Light Magnetic Dipole-dipole Dark Matter}",
    eprint = "2609.06756",
    archivePrefix = "arXiv",
    primaryClass = "hep-ph",
    month = "9",
    year = "2026"
}

@article{DiMauro:2026dqp,
    author = "Di Mauro, Mattia and Shaikh, Halim",
    title = "{Solar Capture Tests of Inelastic Dark Matter after the LZ High-Recoil Event}",
    eprint = "2609.06760",
    archivePrefix = "arXiv",
    primaryClass = "hep-ph",
    month = "9",
    year = "2026"
}

@article{Alhazmi:2026efz,
    author = "Alhazmi, Haider and Kim, Doojin and Kong, Kyoungchul and Park, Jong-Chul and Shin, Seodong",
    title = "{High-Energy Nuclear Recoils from Boosted Dark Matter for the LZ 248-keV Event: Beyond the Halo-Dependent High-Velocity Tail}",
    eprint = "2609.06890",
    archivePrefix = "arXiv",
    primaryClass = "hep-ph",
    month = "9",
    year = "2026"
}

@article{Ahmed:2026qjg,
    author = "Ahmed, Waqas and Leontaris, George K.",
    title = "{A Dark-Dimension Origin of Geometric Inelastic Dark Matter: The LUX-ZEPLIN High-Recoil Event and Multi-Target Tests}",
    eprint = "2609.07138",
    archivePrefix = "arXiv",
    primaryClass = "hep-ph",
    month = "9",
    year = "2026"
}

@article{Bandyopadhyay:2026gjw,
    author = "Bandyopadhyay, Disha and Borah, Debasish and Borah, Pankaj",
    title = "{LZ nuclear recoil event from inelastic singlet-doublet scalar dark matter}",
    eprint = "2609.07451",
    archivePrefix = "arXiv",
    primaryClass = "hep-ph",
    month = "9",
    year = "2026"
}

@article{Kannike:2026qyl,
    author = "Kannike, Kristjan and Raidal, Martti and Strumia, Alessandro",
    title = "{Boosted dark particles and the LZ nuclear recoil event}",
    eprint = "2609.07742",
    archivePrefix = "arXiv",
    primaryClass = "hep-ph",
    month = "9",
    year = "2026"
}

@article{Borah:2026zwf,
    author = "Borah, Debasish and Sahoo, Sujit Kumar and Sahu, Narendra and Sharma, Shashwat",
    title = "{Inelastic Singlet-Doublet Fermion Dark Matter in light of the 248 keV LZ event}",
    eprint = "2609.07800",
    archivePrefix = "arXiv",
    primaryClass = "hep-ph",
    month = "9",
    year = "2026"
}

@article{Bose:2026ndd,
    author = "Bose, Debajit and others",
    title = "{Not so good $\nu$s for Higgsino dark matter as LZ excess: stringent limits from Super-Kamiokande and IceCube}",
    eprint = "2609.07807",
    archivePrefix = "arXiv",
    primaryClass = "hep-ph",
    month = "9",
    year = "2026"
}

@article{Bisal:2026khf,
    author = "Bisal, Subhadip and Cao, Junjie and Li, Fei",
    title = "{Higgsino Dark Matter Interpretation of the LZ High-Recoil Event in the GNMSSM with TeV-Scale Gauginos}",
    eprint = "2609.07811",
    archivePrefix = "arXiv",
    primaryClass = "hep-ph",
    month = "9",
    year = "2026"
}

@article{Khan:2026nwp,
    author = "Khan, Imtiaz and Capozziello, Salvatore and Mustafa, G. and Atamurotov, Farruh and Abdujabbarov, Ahmadjon and Yuan, Chengxun",
    title = "{Nuclear interference versus dark sector excitation in the 248 keV LUX-ZEPLIN recoil candidate}",
    eprint = "2609.09230",
    archivePrefix = "arXiv",
    primaryClass = "hep-ph",
    month = "9",
    year = "2026"
}

@article{Yuan:2026djt,
    author = "Yuan, Guan-Wen and Zhang, Bo and Cao, Wen-Yu and Feng, Lei and Yang, Ruizhi",
    title = "{ALP-mediated inelastic dark matter and the LUX-ZEPLIN high-recoil candidate event LZ230616}",
    eprint = "2609.08893",
    archivePrefix = "arXiv",
    primaryClass = "hep-ph",
    month = "9",
    year = "2026"
}

@article{Zhu:2026dag,
    author = "Zhu, Pengxuan and Dalla Valle Garcia, Giovani and Wang, Xuan-Gong and Thomas, Anthony W. and White, Martin J.",
    title = "{Endothermic dark matter with a light dark photon and the LUX--ZEPLIN high-energy nuclear-recoil candidate}",
    eprint = "2609.09015",
    archivePrefix = "arXiv",
    primaryClass = "hep-ph",
    month = "9",
    year = "2026"
}

@article{Asadi:2026iot,
    author = "Asadi, Pouya and Batz, Austin and Fox, Patrick J. and Homiller, Samuel D. and Kribs, Graham D.",
    title = "{For Whom the Xenon Recoils: Magnetic Inelastic Dark Baryons}",
    eprint = "2609.09107",
    archivePrefix = "arXiv",
    primaryClass = "hep-ph",
    reportNumber = "PITT-PACC-2614, FERMILAB-PUB-26-0660-T",
    month = "9",
    year = "2026"
}

@article{Lee:2026xxh,
    author = "Lee, Vincent S. H. and Randall, Lisa",
    title = "{A Warped Extra Dimensional Candidate for the LZ 248 keV Event}",
    eprint = "2609.09136",
    archivePrefix = "arXiv",
    primaryClass = "hep-ph",
    reportNumber = "N3AS-26-020",
    month = "9",
    year = "2026"
}

@article{Lee:2026jxl,
    author = "Lee, Seung J. and Youn, Taewook",
    title = "{Mixing-suppressed inelastic dark matter: a minimal model for the LZ 248 keV event}",
    eprint = "2609.09138",
    archivePrefix = "arXiv",
    primaryClass = "hep-ph",
    month = "9",
    year = "2026"
}

@article{Langhoff:2026ujr,
    author = "Langhoff, Kevin",
    title = "{Heavy Higgsino Interpretation of the LZ Event}",
    eprint = "2609.09385",
    archivePrefix = "arXiv",
    primaryClass = "hep-ph",
    month = "9",
    year = "2026"
}

@article{Chatterjee:2026scv,
    author = "Chatterjee, Arindam and Das, Debottam and Pasha, Syed Adil and Pukhov, Alexander and Puri, Rahul",
    title = "{Radiative Corrections to the Direct Detection of Inelastic Scattering of Higgsino-like Neutralino Dark Matter}",
    eprint = "2609.09830",
    archivePrefix = "arXiv",
    primaryClass = "hep-ph",
    month = "9",
    year = "2026"
}

@article{He:2026hqz,
    author = "He, Yuxuan",
    title = "{Transition magnetic-dipole dark matter and the LZ230616 high-recoil candidate}",
    eprint = "2609.10453",
    archivePrefix = "arXiv",
    primaryClass = "hep-ph",
    month = "9",
    year = "2026"
}

@article{Fan:2026hzw,
    author = "Fan, Zi-Tong and He, Hong-Jian and Wang, Yu-Chen and Zhao, Yue",
    title = "{Inelastic Dark Matter and High-Energy Recoil Signatures in LZ}",
    eprint = "2609.10491",
    archivePrefix = "arXiv",
    primaryClass = "hep-ph",
    month = "9",
    year = "2026"
}

@article{Qi:2026vyp,
    author = "Qi, XinXin and Sun, Hao",
    title = "{Nonthermal Solar Stalling of an Inelastic Scalar Signal in Xenon}",
    eprint = "2609.10636",
    archivePrefix = "arXiv",
    primaryClass = "hep-ph",
    month = "9",
    year = "2026"
}

@article{Li:2026fci,
    author = "Li, Qinxun and Liu, Shengyi",
    title = "{Implications of Inelastic Dark Matter for Primordial Dark-Star Evolution: Kinematic Thresholds and Nonthermal Capture}",
    eprint = "2609.10662",
    archivePrefix = "arXiv",
    primaryClass = "hep-ph",
    month = "9",
    year = "2026"
}

@article{Frolovsky:2026tvq,
    author = "Frolovsky, Daniel and Ketov, Sergei V.",
    title = "{Higgsino dark matter in the Starobinsky supergravity with the MSSM in light of the LUX-ZEPLIN event}",
    eprint = "2609.11241",
    archivePrefix = "arXiv",
    primaryClass = "hep-ph",
    reportNumber = "IPMU26-0034",
    month = "9",
    year = "2026"
}

@article{Heikinheimo:2026kwp,
    author = "Heikinheimo, Matti and Zimmermann, Niklas",
    title = "{Cosmic ray boosted dark matter with momentum dependent interactions can explain the LZ 248 keV event}",
    eprint = "2609.11600",
    archivePrefix = "arXiv",
    primaryClass = "hep-ph",
    month = "9",
    year = "2026"
}

@article{Nguyen:2026lui,
    author = "Nguyen, Thong T. Q. and Linden, Tim and Hooper, Dan",
    title = "{Solar Neutrino Constraints on Inelastic Dark Matter Scattering in Light of Recent LUX-ZEPLIN Observations}",
    eprint = "2609.11833",
    archivePrefix = "arXiv",
    primaryClass = "hep-ph",
    month = "9",
    year = "2026"
}

@article{Okada:2026upm,
    author = "Okada, Hiroshi and Shigekami, Yoshihiro and Wu, Jia-Jun",
    title = "{Can a minimal radiative seesaw explain the LZ 248 keV event?}",
    eprint = "2609.13038",
    archivePrefix = "arXiv",
    primaryClass = "hep-ph",
    month = "9",
    year = "2026"
}

@article{Mahapatra:2026glu,
    author = "Mahapatra, Satyabrata and Paul, Partha Kumar",
    title = "{Boosted or Inelastic? Discriminating Interpretations of the LZ 248 keV Event}",
    eprint = "2609.14799",
    archivePrefix = "arXiv",
    primaryClass = "hep-ph",
    month = "9",
    year = "2026"
}

@article{Borah:2026ris,
    author = "Borah, Pankaj and Mahapatra, Satyabrata and Nath, Newton and Paul, Partha Kumar",
    title = "{Inelastic Dark Matter at LZ from Radiative Dirac Neutrino Mass Paradigm}",
    eprint = "2609.15027",
    archivePrefix = "arXiv",
    primaryClass = "hep-ph",
    month = "9",
    year = "2026"
}

@article{Barman:2026omh,
    author = "Barman, Basabendu",
    title = "{Did LZ see modified gravity?}",
    eprint = "2609.15118",
    archivePrefix = "arXiv",
    primaryClass = "hep-ph",
    month = "9",
    year = "2026"
}

@article{Ghosh:2026txe,
    author = "Ghosh, Aditya and Chavez, Ilumi and Kelso, Chris",
    title = "{Confronting the Higgsino Interpretation of the LZ Event with Astrophysical Uncertainties and the Solar Capture Constraints}",
    eprint = "2609.15321",
    archivePrefix = "arXiv",
    primaryClass = "hep-ph",
    month = "9",
    year = "2026"
}

@article{He:2026idw,
    author = "He, Xiao-Gang and Hong, Xuan and Jeesun, Sk",
    title = "{Hadrophilic inelastic freeze-in dark matter in $q_1-q_2$ gauge extension and the high energy LZ event}",
    eprint = "2609.15714",
    archivePrefix = "arXiv",
    primaryClass = "hep-ph",
    month = "9",
    year = "2026"
}

@article{Lian:2026hpm,
    author = "Lian, Jingwei and Yang, Jin Min",
    title = "{Explain the LZ High-Energy Recoil Event with Inelastic Sneutrino Dark Matter in Supersymmetry}",
    eprint = "2609.15742",
    archivePrefix = "arXiv",
    primaryClass = "hep-ph",
    month = "9",
    year = "2026"
}

@article{Baer:2026yrt,
    author = "Baer, Howard and Barger, Vernon",
    title = "{Argon as the test of two interpretations of the LZ 248 keV recoil}",
    eprint = "2609.15782",
    archivePrefix = "arXiv",
    primaryClass = "hep-ph",
    month = "9",
    year = "2026"
}

@article{Uttayarat:2026isp,
    author = "Uttayarat, P. and Julio, J. and Primulando, R.",
    title = "{DM induced neutron disappearance as the origin of the LZ nuclear recoil event}",
    eprint = "2609.15933",
    archivePrefix = "arXiv",
    primaryClass = "hep-ph",
    month = "9",
    year = "2026"
}

@article{Palmisano:2026kuj,
    author = "Palmisano, Stefano and Tammaro, Michele and Tesi, Andrea",
    title = "{Inferring dark matter masses and interactions from high recoil energy events in LUX-ZEPLIN}",
    eprint = "2609.15985",
    archivePrefix = "arXiv",
    primaryClass = "hep-ph",
    month = "9",
    year = "2026"
}

@article{DiMauro:2026ymt,
    author = "Di Mauro, Mattia",
    title = "{Testing Higgs-Coupled Minimal Dark Matter with Solar Neutrinos after the LZ High-Recoil Event}",
    eprint = "2609.19174",
    archivePrefix = "arXiv",
    primaryClass = "hep-ph",
    month = "9",
    year = "2026"
}

@article{An:2026pkc,
    author = "An, Haipeng and Gao, Fei and Liu, Jia and Liu, Minghao and Xu, Changlong",
    title = "{Cosmological Constrained Axion-Portal Inelastic Dark Matter for the LZ Event}",
    eprint = "2609.17412",
    archivePrefix = "arXiv",
    primaryClass = "hep-ph",
    month = "9",
    year = "2026"
}

@article{Xing:2026civ,
    author = "Xing, Chuan-Yang",
    title = "{Galactic Endothermic Production and Exothermic Detection of Excited Dark Matter: Implications for LUX-ZEPLIN}",
    eprint = "2609.17935",
    archivePrefix = "arXiv",
    primaryClass = "hep-ph",
    month = "9",
    year = "2026"
}

@article{Nagata:2026pbj,
    author = "Nagata, Natsumi and Yanagida, Tsutomu T.",
    title = "{Asymmetric Inelastic Dark Matter and the LUX-ZEPLIN event}",
    eprint = "2609.18564",
    archivePrefix = "arXiv",
    primaryClass = "hep-ph",
    month = "9",
    year = "2026"
}

@article{Schwaller:2013baa,
    author = "Schwaller, Pedro and Zurita, Jose",
    title = "{Compressed electroweakino spectra at the LHC}",
    eprint = "1312.7350",
    archivePrefix = "arXiv",
    primaryClass = "hep-ph",
    reportNumber = "CERN-PH-TH-2013-319, MITP-13-084",
    doi = "10.1007/JHEP03(2014)060",
    journal = "JHEP",
    volume = "03",
    pages = "060",
    year = "2014"
}

@article{Mahbubani:2017gjh,
    author = "Mahbubani, Rakhi and Schwaller, Pedro and Zurita, Jose",
    title = "{Closing the window for compressed Dark Sectors with disappearing charged tracks}",
    eprint = "1703.05327",
    archivePrefix = "arXiv",
    primaryClass = "hep-ph",
    reportNumber = "CERN-TH-2017-054, TTP17-005, MITP/17-011",
    doi = "10.1007/JHEP06(2017)119",
    journal = "JHEP",
    volume = "06",
    pages = "119",
    year = "2017",
    note = "[Erratum: JHEP 10, 061 (2017)]"
}

@article{Capdevilla:2021fmj,
    author = "Capdevilla, Rodolfo and Meloni, Federico and Simoniello, Rosa and Zurita, Jose",
    title = "{Hunting wino and higgsino dark matter at the muon collider with disappearing tracks}",
    eprint = "2102.11292",
    archivePrefix = "arXiv",
    primaryClass = "hep-ph",
    reportNumber = "DESY-21-019, IFIC/21-03",
    doi = "10.1007/JHEP06(2021)133",
    journal = "JHEP",
    volume = "06",
    pages = "133",
    year = "2021"
}

@article{ATLAS:2026hnb,
    author = "Aad, Georges and others",
    collaboration = "ATLAS",
    title = "{Search for long-lived charginos and {\ensuremath{\tau}}-sleptons using final states with a disappearing track in pp collisions at $ \sqrt{s}=13 $ TeV with the ATLAS detector}",
    eprint = "2603.08315",
    archivePrefix = "arXiv",
    primaryClass = "hep-ex",
    reportNumber = "CERN-EP-2026-044",
    doi = "10.1007/JHEP07(2026)152",
    journal = "JHEP",
    volume = "07",
    pages = "152",
    year = "2026"
}

@article{CMS:2026ias,
    author = "Hayrapetyan, Aram and others",
    collaboration = "CMS",
    title = "{Search for electroweakinos in compressed-spectrum scenarios with low-momentum isolated tracks in proton-proton collisions at $\sqrt{s}$ = 13 TeV}",
    eprint = "2604.25604",
    archivePrefix = "arXiv",
    primaryClass = "hep-ex",
    reportNumber = "CMS-SUS-24-012, CERN-EP-2026-103",
    month = "4",
    year = "2026"
}

@article{Planck:2018vyg,
    author = "Aghanim, N. and others",
    collaboration = "Planck",
    title = "{Planck 2018 results. VI. Cosmological parameters}",
    eprint = "1807.06209",
    archivePrefix = "arXiv",
    primaryClass = "astro-ph.CO",
    doi = "10.1051/0004-6361/201833910",
    journal = "Astron. Astrophys.",
    volume = "641",
    pages = "A6",
    year = "2020",
    note = "[Erratum: Astron.Astrophys. 652, C4 (2021)]"
}

@article{Alguero:2023zol,
    author = "Alguero, G. and Belanger, G. and Boudjema, F. and Chakraborti, S. and Goudelis, A. and Kraml, S. and Mjallal, A. and Pukhov, A.",
    title = "{micrOMEGAs 6.0: N-component dark matter}",
    eprint = "2312.14894",
    archivePrefix = "arXiv",
    primaryClass = "hep-ph",
    doi = "10.1016/j.cpc.2024.109133",
    journal = "Comput. Phys. Commun.",
    volume = "299",
    pages = "109133",
    year = "2024"
}

@article{Belanger:2026asz,
    author = "Belanger, G. and Belyaev, A. and Bernal, N. and Boudjema, F. and Chakraborti, S. and Goudelis, A. and Pukhov, A.",
    title = "{micrOMEGAs 7: Beyond standard cosmology}",
    eprint = "2606.06645",
    archivePrefix = "arXiv",
    primaryClass = "hep-ph",
    reportNumber = "IPPP/26/46",
    month = "6",
    year = "2026"
}

@article{Becker:2026icc,
    author = "Becker, Mathias and Copello, Emanuele and Harz, Julia and Napetschnig, Martin",
    title = "{Sommerfeld Effect and Bound State Formation for Dark Matter Models with Colored Mediators with SE+BSF4DM}",
    eprint = "2601.03026",
    archivePrefix = "arXiv",
    primaryClass = "hep-ph",
    reportNumber = "TUM-HEP-1571/25, MITP-25-058",
    month = "1",
    year = "2026"
}

@article{Becker:2025vgq,
    author = "Becker, Mathias and Copello, Emanuele and Harz, Julia and Napetschnig, Martin",
    title = "{Manual for SE+BSF4DM -- A micrOMEGAs package for Sommerfeld Effect and Bound State Formation in colored Dark Sectors}",
    eprint = "2512.02155",
    archivePrefix = "arXiv",
    primaryClass = "hep-ph",
    reportNumber = "TUM-HEP-1572/25, MITP-25-077",
    month = "12",
    year = "2025"
}

@article{Maier:2026qpr,
    author = "Maier, Benedikt and Spannowsky, Michael",
    title = "{The LHC is not enough: the LZ High-Recoil Event at FCC-hh and a Muon Collider}",
    eprint = "2609.26870",
    archivePrefix = "arXiv",
    primaryClass = "hep-ph",
    month = "9",
    year = "2026"
}

@misc{LHCSUSYXS:Wino14,
    author = "{LHC SUSY Cross Section Working Group}",
    title = "{NLO+NNLL wino-like chargino--neutralino cross sections at 14 TeV}",
    howpublished = "{\href{https://twiki.cern.ch/twiki/bin/view/LHCPhysics/SUSYNLONNLLCrossSections14TeVWinoN2C1}{LHC SUSY Cross Section Working Group online table}}",
    year = "2019"
}

@article{Roy:2026wzb,
    author = "Roy, Subhojit and Schwaller, Pedro and Wagner, Carlos E. M.",
    title = "{Interpreting the High-Recoil LUX-ZEPLIN Event with Bino-/Singlino-like and Higgsino Dark Matter}",
    eprint = "2609.40234",
    archivePrefix = "arXiv",
    primaryClass = "hep-ph",
    month = "9",
    year = "2026"
}

@article{Bozorgnia:2026emp,
    author = "Bozorgnia, Nassim and Contreras, Samantha and Gelmini, Graciela B. and Kamaha, Alvine C. and Reynoso-Cordova, Javier and Xu, Yongheng",
    title = "{LMC-Perturbed LZ Dark Matter Landscape}",
    eprint = "2609.36184",
    archivePrefix = "arXiv",
    primaryClass = "hep-ph",
    month = "9",
    year = "2026"
}

@article{OHare:2026nqi,
    author = "O'Hare, Ciaran A. J.",
    title = "{The high-velocity dark matter halo of the Milky Way in light of the LZ 248 keV event}",
    eprint = "2609.21444",
    archivePrefix = "arXiv",
    primaryClass = "hep-ph",
    month = "9",
    year = "2026"
}

@article{Luengas:2026tlf,
    author = "Luengas, Juan P. and Pospelov, Maxim and Ramani, Harikrishnan",
    title = "{Constraints on Complex Electroweak WIMPs from Solar Capture}",
    eprint = "2609.38316",
    archivePrefix = "arXiv",
    primaryClass = "hep-ph",
    month = "9",
    year = "2026"
}

@article{Das:2026jsq,
    author = "Das, Anirban and Majumdar, Subhabrata and Sen, Manibrata and Srivastav, Amogh",
    title = "{GAIA meets LZ: a high velocity tail-tale sign for dark matter}",
    eprint = "2609.35737",
    archivePrefix = "arXiv",
    primaryClass = "hep-ph",
    month = "9",
    year = "2026"
}

@article{Girmohanta:2026qzo,
    author = "Girmohanta, Sudhakantha and Jung, Tae Hyun",
    title = "{Environment-dependent mass splitting to suppress solar capture in the inelastic-doublet interpretation of the LZ event}",
    eprint = "2609.28129",
    archivePrefix = "arXiv",
    primaryClass = "hep-ph",
    month = "9",
    year = "2026"
}

@article{Khan:2026zuj,
    author = "Khan, Imtiaz and Muhammad, Ali and Mustafa, G. and Atamurotov, Farruh and Abdujabbarov, Ahmadjon and Khan, Mussawir",
    title = "{LZ-Motivated Pseudo-Dirac Higgsinos in the Supersymmetric 331 Model from the Supersymmetric {SU(6)} GUT Model}",
    eprint = "2609.23691",
    archivePrefix = "arXiv",
    primaryClass = "hep-ph",
    month = "9",
    year = "2026"
}

@article{Okada:2026fef,
    author = "Okada, Nobuchika and Raut, Digesh",
    title = "{Endothermic Z'-Portal Dark Matter: LZ-LHC Complementarity}",
    eprint = "2609.21011",
    archivePrefix = "arXiv",
    primaryClass = "hep-ph",
    month = "9",
    year = "2026"
}

\end{document}